\documentclass[aps,PRL,twocolumn,nofootinbib,groupedaddress,superscriptaddress
,longbibliography]{revtex4-2}

\usepackage{amsmath,amssymb}
\usepackage{graphicx}
\usepackage{multirow}
\usepackage{bm}
\usepackage{mathtools}
\usepackage{dsfont}
\usepackage{amsfonts}
\usepackage{xcolor}
\usepackage[normalem]{ulem}
\usepackage{url}
\usepackage{wasysym}
\usepackage{bbm}

\usepackage[colorlinks=true,linkcolor=magenta,citecolor=magenta,hypertexnames=false]{hyperref}
\newcommand{\nocontentsline}[3]{}
\newcommand{\tocless}[2]{\bgroup\let\addcontentsline=\nocontentsline#1{#2}\egroup}

\def\ba#1\ea{\begin{align}#1\end{align}}
\def\bg#1\eg{\begin{gather}#1\end{gather}}
\def\bpm{\begin{pmatrix}}
\def\epm{\end{pmatrix}}

\newcommand{\ket}[1]{|#1\rangle}
\newcommand{\bra}[1]{\langle#1|}

\newcommand{\magenta}[1]{\textcolor{magenta}{#1}}

\newcommand{\ourtitle}{
Dipole-Obstructed Cooper Pairing: Theory and Application to $j=3/2$ Superconductors}

\allowdisplaybreaks

\begin{document}
\title{\textbf{\ourtitle}}

\author{Penghao Zhu} 
\email{zhu.3711@osu.edu}
\affiliation{Department of Physics, The Ohio State University, Columbus, OH 43210, USA}
\author{Rui-Xing Zhang}
\email{ruixing@utk.edu}
\affiliation{Department of Physics and Astronomy, The University of Tennessee, Knoxville, TN 37996, USA}
\affiliation{Department of Materials Science and Engineering, The University of Tennessee, Knoxville, TN 37996, USA}

\begin{abstract}
Like electrons, Cooper pairs can carry a monopole charge if the pairing electrons come from two or more Fermi surfaces with different Chern numbers. In such an instance, a superconductor is necessarily nodal due to an inherent topological pairing obstruction. In this work, we show that a similar obstruction is also possible when there is only one Fermi surface involved in the pairing process. By developing a Chern-vorticity theorem, we have identified a class of Fermi surfaces with a quantized dipolar Berry flux pattern, where all intra-Fermi-surface Cooper pairings are ``dipole-obstructed" and nodal. As a real-world application, we find that the dipole obstruction plays a crucial role in stabilizing the superconducting nodal structure for $j=3/2$ half-Heusler compounds.       
\end{abstract}

\maketitle

\let\oldaddcontentsline\addcontentsline
\renewcommand{\addcontentsline}[3]{}

\magenta{\it Introduction.|} Superconductors (SCs) are generally classified by the symmetry pattern of their Cooper pairs. By definition, electron pairing in a conventional $s$-wave SC is spatially isotropic, generating a uniform energy gap of the normal-state Fermi surface (FS)~\cite{BCS1957}. Meanwhile, an unconventional SC such as the cuprates can feature a gapless spectrum when its anisotropic pairing order $\Delta({\bf k})$ has symmetry-enforced zeros in the momentum space~\cite{sigrist1991RMP,tsuei2000RMP,kim2018beyond}. In experiments, the SC gap structure can be feasibly probed by angle-resolved photoemission spectroscopy~\cite{damascelli2003ARPES}, scanning tunneling spectroscopy~\cite{fischer2007STMRMP}, penetration depth measurements~\cite{prozorov2006penetration}, etc. Such gap information often offers valuable insights into unraveling the nature of Cooper pairs for a new SC candidate.  

The pairing symmetry, however, does not fully account for the gap structure of SCs. For example, a SC is found to be necessarily nodal, once the electrons forming a Cooper pair come from two FSs with opposite Berry monopole charges~\cite{murakami2003berry, Li2018topological, wang2016densitywave,Munoz2020monopole}. Such an inability to develop a full energy gap is intrinsic to the topological texture of the FS, which holds even when $\Delta({\bf k})$ is constant. This striking phenomenon has sparked a growing research interest in uncovering similar mechanisms of topologically obstructed nodal pairing orders for FSs carrying an Euler index~\cite{Yu2022euler,yu2023euler,wang2024molecular} or a $\mathbb{Z}_2$ index~\cite{Sun2020mathbb}. Notably, most existing theories have assumed the obstructed pairings to involve multiple FSs, while generalizations to intra-FS electron pairings are less explored.

In this work, we have identified a new class of FSs where all intra-FS Cooper pairings exhibit topology-enforced zeros. The FS of our interest encloses a quantized dipole of the Berry curvature~\cite{Nelson2022delicate}, i.e., there exists a Berry flux of $\pm 2\pi$ through either half of the FS, while the net Berry flux vanishes. As a proof of concept, we consider a minimal model with such a Berry-dipole FS and find all intra-FS Cooper pairings are {\it dipole-obstructed} from fully gapping out the FS. Specifically, the resulting SC state always exhibits zero-energy Weyl nodes and/or nodal loops in the Bogoliubov-de Gennes (BdG) spectrum. Finally, we revisit the half-Heusler SCs such as YPtBi~\cite{butch2011SC,kim2018beyond} and LuPdBi~\cite{nakajima2015topological,pavlosiuk2015shubnikov,tuning2021ishihara}, which are believed to feature a mixed-parity singlet-septet pairing~\cite{pairing2016brydon}. We find that the dipole obstruction naturally exists in the septet pairing channel and further clarify its contribution to the BdG line nodes observed in experiments.

\begin{figure}[t]
    \centering
    \includegraphics[width=0.95\columnwidth]{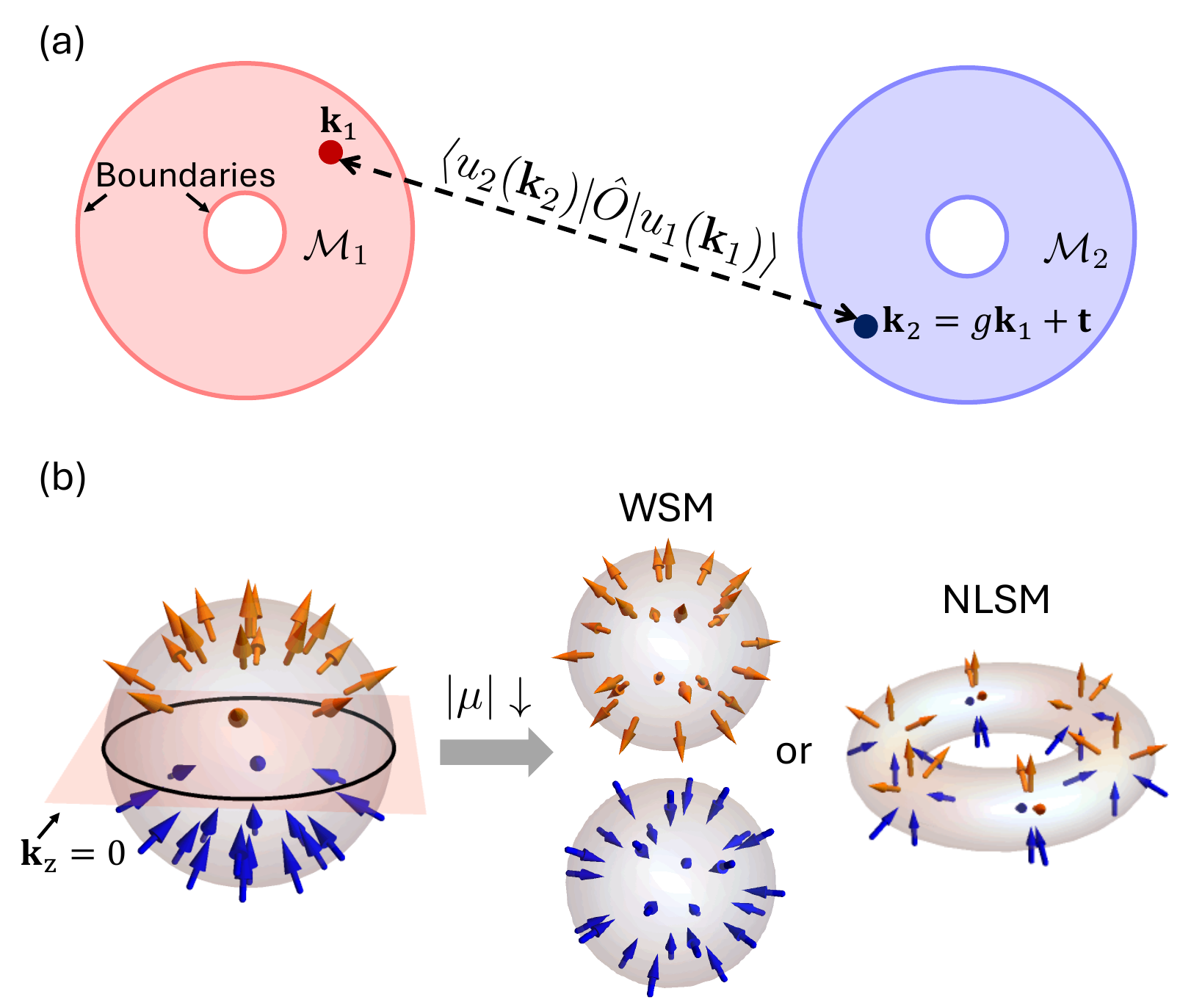}
    \caption{(a) The Chern-vorticity theorem informs the phase vortex of $\hat{O}$ when projecting onto two closed or effectively closed manifolds ${\cal M}_{1,2}$. (b) Berry-dipole FSs for $h_0$, where the arrows denote the Berry curvature vectors on the FS. A change of $\mu$ or $\Sigma$ can modify the FS topology, while the quantization of Berry dipole always remains robust.}
    \label{fig:illustration}
\end{figure}

\magenta{\it Chern-Vorticity Theorem and Pairing Zeros.|} We start by presenting a Chern-vorticity theorem that will guide us to the target FSs. As shown in Fig.~\ref{fig:illustration}(a), we consider two 2D closed or \textit{effectively} closed manifolds $\mathcal{M}_{1}$ and $\mathcal{M}_{2}$ in ${\bf k}$-space, as well as two Bloch states $\ket{\psi_{1}(\mathbf{k}_1)}$ and $\ket{\psi_{2}(\mathbf{k}_2)}$ with ${\bf k}_i\in {\cal M}_i$. Here, we define an open manifold to be effectively closed if the Bloch states over each of its boundaries are identical. Generally, $\mathcal{M}_{1}$ and $\mathcal{M}_{2}$ are related by a $\mathbf{k}$-space transformation $\{g|{\bf t}\}$ with $\mathbf{k}_{2}=g\mathbf{k}_{1}+\mathbf{t}.$ Here $g$ denotes a point-group operation (e.g., rotation and mirror) that satisfies $g^{-1}=g^{T}\in \mathbb{R}$ and $\mathbf{t}$ is a translation in $\mathbf{k}$-space. We now consider the matrix element of a general two-particle operator $\hat{O}$ bridging electrons on $\mathcal{M}_{1}$ and $\mathcal{M}_{2}$:
\begin{equation}
    {\cal O}({\bf k}_1) = \bra{\psi_{1}(\mathbf{k}_1)}\hat{O}\ket{\psi_{2}(\mathbf{k}_{2})} = |{\cal O}({\bf k}_1)|e^{i \varphi({\bf k}_1)}.
    \label{eq:projected operator}
\end{equation}
When $\varphi$ displays a vortex pattern around ${\bf k}_1={\bf k}_{v_i}$, the value of ${\cal O}({\bf k}_{v_i})$ necessarily vanishes and is thus topologically obstructed. Such a vortex $v_i$ is captured by a vorticity index around ${\bf k}_{v_i}$, defined  as $\nu_i=1/(2\pi)\oint d\mathbf{k}_{1}\cdot\partial_{\mathbf{k}_{1}}\varphi({\bf k}_1)\in\mathbb{Z}$. The Chern-vorticity theorem proven in the Supplemental Material (SM)~\cite{supp} dictates that the net vorticity on ${\cal M}_1$, $\nu = \sum_{i} \nu_i \in\mathbb{Z}$, is determined by  
\begin{eqnarray}
    \nu &=& {\cal C}_{1}-({\operatorname{det}g}) {\cal C}_{2}+ {\cal I_{\varphi}}, \nonumber \\
    {\cal I}_{\varphi} &=& \sum_{\partial {\cal M}_1} \oint_{\partial {\cal M}_1}\frac{d\mathbf{k}_{1}}{2\pi}\cdot\partial_{\mathbf{k}_{1}}\varphi(\mathbf{k}_1),
    \label{eq:Chern-vort}
\end{eqnarray}
where ${\cal C}_{\alpha}$ is the single-particle Chern number of $\ket{u_{\alpha}(\mathbf{k}_{\alpha})} \equiv e^{-i{\bf k}_\alpha \cdot {\bf r}_\alpha}\ket{\psi_{2}(\mathbf{k}_\alpha)}$ on ${\cal M}_\alpha$. The loop integral ${\cal I}_\varphi$ is summed over all possible boundaries of ${\cal M}_1$, which, by default, vanishes for a closed ${\cal M}_1$.

Focusing on SCs, we choose $\ket{\psi_{1}(\mathbf{k}_{1})}=\ket{\xi^{(e)}_{1}(\mathbf{k}_{1})}$ and $\ket{\psi_{2}(\mathbf{k}_{2})}=\ket{\xi^{(e)\star}_{2}(\mathbf{k}_{2})}$ in order to properly project the pairing operator $\hat{O}$, since $\hat{O}$ physically describes hoppings between electrons and holes. Here, $\ket{\xi^{(e)}_{\alpha}(\mathbf{k}_{\alpha})}$ is the Bloch state of electrons on the Fermi surface ${\cal M}_{\alpha}$. It is crucial to note that ${\cal C}_2$, the Chern number for $\ket{\psi_{2}(\mathbf{k}_{2})}$, is exactly opposite to ${\cal C}_2^{(e)}$, the electronic Chern number for $\ket{\xi^{(e)}_{2}(\mathbf{k}_{2})}$ on ${\cal M}_2$~\cite{supp}. As a concrete example, we immediately arrive at $\nu=2$ when (i) $\mathcal{M}_{1,2}$ are both closed FSs with ${\cal C}_1 = {\cal C}_2 = +1$ and (ii) $g$ is the spatial inversion with ${\operatorname{det}g} = -1$, which exactly corresponds to the obstructed nodal pairing for monopole SCs in Ref.~\cite{Li2018topological}. Therefore, the Chern-vorticity theorem manifests as a natural generalization of Ref.~\cite{Li2018topological}, but its application transcends the realm of SC systems with inter-FS pairings~\cite{bobrow2020monopole,bultinck2020mechanism,zhu2023anomalous}.   

\magenta{\it Berry-Dipole Fermi Surface.|} The above Chern-vorticity theorem suggests that for a single Fermi surface ${\cal M} \equiv {\cal M}_+\cup {\cal M}_-$, the intra-FS pairing between two patches ${\cal M}_{\pm}$ can be topologically obstructed if each patch is effectively closed and carries a nonzero quantized Berry flux. Further requiring the net Berry flux to be zero, we are thus looking for a FS with a quantized dipolar texture of Berry curvature, i.e., a Berry-dipole FS~\cite{Sun2018conversion,Nelson2022delicate,graf2023massless,zhuang2024berry,tyner2024dipolar}.  

As discussed in Refs.~\cite{Sun2018conversion, Nelson2022delicate} and reviewed in the SM~\cite{supp}, quantization of a Berry-dipole can be achieved by enforcing a mirror symmetry along the dipole axis. As a concrete example, we consider a minimal model with Berry-dipole physics proposed in Ref.~\cite{Nelson2022delicate}, 
\begin{equation}
\label{eq:berrydipole}
h_{0} =2k_z (k_x \sigma_x + k_y \sigma_y) + (k_{x}^2+k_{y}^2-k_{z}^2+\Sigma) \sigma_z.
\end{equation}
Besides mirror symmetry along $z$-direction $M_z=i\sigma_z$, $h_0$ respects a continuous rotation symmetry around $z$ with $C_\theta = \text{exp}[-i J_z \theta]$, where $J_z = \text{diag}(3/2, 1/2)$ is a diagonal matrix. Regularizing $h_0$ will reduce the symmetry group to $C_{4h}$, which will be exploited for later pairing analysis. When $\Sigma>0$, $h_{0}$ describes a 3D minimal Weyl semimetal (WSM) with two Weyl nodes at $k_z=\pm\sqrt{\Sigma}$, respectively. At $\Sigma=0$, the Weyl nodes merge at $\Gamma$ to form a quadratic band touching. The annihilation of Weyl nodes, however, does not lead to a mass generation for the low-energy electrons. Instead, we note that $h_{0}$ with $\Sigma<0$ features a 1D doubly degenerate nodal loop in the $k_{z}=0$ plane, i.e., a nodal-loop semimetal (NLSM). This unexpected robustness of Weyl nodes arises from the Berry-dipole {\it delicate} topological charges~\cite{Sun2018conversion,Nelson2022delicate}. 

Let us generally denote the FS patch with $k_{z}>0$ ($k_{z}<0$) as $\mathcal{M}_{+}$ ($\mathcal{M}_{-}$). For $|\mu|>|\Sigma|$, we always find a single spherical-like FS ${\cal M}$ regardless of the value of $\Sigma$, where $\mathcal{M}_{+}$ and $\mathcal{M}_{-}$ are hence the north and south hemispheres, respectively. Undergoing the Lifshitz transition with $|\mu|<|\Sigma|$, the WSM with $\Sigma>0$ comprises a pair of closed spherical FSs (i.e., ${\cal M}_\pm$), while a single torus-like FS is found for the NLSM ($\Sigma<0$) with $\mathcal{M}_{\pm}$ now being both open and of an annulus shape. Evidently, the boundaries of $\mathcal{M}_{+}$ and $\mathcal{M}_{-}$, if present, are always sitting in the $k_{z}=0$ plane regardless of $\Sigma$. Consequently, the eigenstates of $h_{0}$ along each boundary can always be uniform, owing to the $M_z$ symmetry. Hence, we conclude that $\mathcal{M}_{+}$ and $\mathcal{M}_{-}$ must be either closed or effectively closed for all choices of $\mu$ and $\Sigma$. This guarantees the FS(s) of $h_0$ to always feature a quantized Berry dipole, as explicitly confirmed in Fig.~\ref{fig:illustration}(b).

\magenta{\it Dipole-Obstructed Pairing.|}  We are now ready to explore Cooper pairing physics on the above Berry-dipole FS by updating $h_0$ to a BdG Hamiltonian,
\begin{equation}
\label{eq:bdgHbd}
H(\mathbf{k})=\begin{pmatrix}
h_{0}(\mathbf{k})-\mu & \Delta (\mathbf{k})
\\
\Delta^{\dag} (\mathbf{k}) & \mu -h^{T}_{0}(-\mathbf{k})
\end{pmatrix}.
\end{equation}
The pairing matrix generally takes the form $\Delta(\mathbf{k})=d_{0}(\mathbf{k})\sigma_{0}+\mathbf{d}(\mathbf{k})\cdot\boldsymbol{\sigma}$ and the Fermi statistics requires $\Delta({\bf k})= - \Delta^T(-{\bf k})$. For our purpose, we consider expanding $d_i({\bf k})$ up to ${\cal O}({\bf k}^2)$ and classify all pairing channels based on the irreducible representations (irreps) of $C_{4h}$ group in Table.~\ref{tab:pairing_class_BD}~\cite{supp}. For $|\mu|>|\Sigma|$, $\Delta({\bf k})$ describes the intra-FS pairings (as there is only one FS). Following Eq.~\ref{eq:projected operator}, the projected Cooper pairing onto ${\cal M}_\pm$ is given by $\Delta_{\text{eff}}=\bra{\xi^{(e)}_{+}(\mathbf{k})}\Delta(\mathbf{k})\ket{\xi^{(e)\star}_{-}(-\mathbf{k})}$, where $\ket{\xi^{(e)}_{\pm}}$ are normal states over $\mathcal{M}_{\pm}$.

We first note that a general pairing with ${\cal I}_\varphi=0$ must be topologically obstructed since $\nu = C_1 + C_2 = 2\text{sgn}(\mu)$, which directly applies to all $\sigma_{x,y}$ pairings. To see this, note that $|\xi_{\pm}^{(e)}(k_z=0)\rangle$ must be eigenstates of $\sigma_{z}$, since $h_0|_{k_z=0}\sim \sigma_z$. As a result, any $\Delta({\bf k})\sim \sigma_{x,y}$ will vanish on the equator upon projection, further leading to ${\cal I}_\varphi = 0 $. Meanwhile, the $\sigma_\pm$ component of $\Delta({\bf k})$ can contribute to ${\cal I}_\varphi$ by the winding phase of $d_{+}$ ($d_{-}$) around the equator for $\mu>0$ ($\mu<0$), so long as $d_\pm (k_z=0) \neq 0$. Here, we have defined $\sigma_\pm = (\sigma_0 \pm \sigma_z)/2$ and $d_\pm ({\bf k}) = d_0 ({\bf k}) \pm d_z ({\bf k})$. Based on Table.~\ref{tab:pairing_class_BD}, it is straightforward to see that $A_{u}$ and $B_{u}$ pairings are the only two pairing channels that can contribute to ${\cal I}_\varphi$. Since $d_\pm({\bf k})$ must be an odd function of $k_\pm=k_x\pm i k_y$ for both $A_u$ and $B_u$ irreps, the corresponding phase winding of $d_{\pm}$ around the equator must be $(4m+1)\pi$ with $m\in\mathbb{Z}$~\cite{supp}. This necessarily leads to a non-zero odd-integer-valued $\nu$ for all relevant $A_u$ and $B_u$ pairings. Since all other pairings feature ${\cal I}_\varphi=0$ and $\nu=2$, we conclude that all intra-FS pairing channels for the Berry-dipole FS are obstructed and nodal. 

\begin{table}[t]
    \centering
    \begin{tabular}{cccccccc} \hline
       $A_g$ & $B_g$ & $^1E_g$  & $^2E_g$ & $A_u$ & $B_u$ & $^1E_u$ & $^2E_u$ \\ \hline
        $(k_x^2 - k_y^2)\sigma_y$ & $\Delta_0 \sigma_y$ & $k_z k_- \sigma_y$ & $k_z k_+ \sigma_y$ & $k_\pm \sigma_\pm$ & $k_\mp \sigma_\pm$ & $k_z \sigma_-$ & $k_z\sigma_+$ \\
         $k_x k_y\sigma_y$ &  &  &  &  & $k_z \sigma_x$ & $k_-\sigma_x$ & $k_+\sigma_x$ \\ \hline
    \end{tabular}
    \caption{Pairing classification for Berry-dipole FS following the $C_{4h}$ irreps. For $B_g$-pairing, $d_y=\Delta_0$ is a constant.}
    \label{tab:pairing_class_BD}
\end{table}

As a concrete example, let us focus on $\Delta = d_x ({\bf k}) \sigma_x$ to explicitly illustrate the dipole-induced obstruction and relegate the discussions on other pairing channels in the SM~\cite{supp}. Without loss of generality, we set $\Sigma=0$ and the FS at $\mu\neq 0$ is a sphere of radius $\sqrt{|\mu|}$, which can be parameterized by a polar angle $\theta \in [0,\pi]$ and an azimuthal angle $\phi \in [0,2\pi)$. For $\mu<0$, we find the electron wavefunction to be $|\xi^{(e,I)}({\bf k})\rangle = (-\cos \theta e^{-i\phi}, \sin\theta )^T$, which is a constant spinor $(0,1)^T$ at the equator. Under this gauge choice, ${\cal M}_\pm$ are both effectively closed, while $|\xi^{(e,I)}({\bf k})\rangle$ becomes singular at the north pole ($\theta=0$). This implies an obstruction to define a globally smooth gauge, thanks to the non-zero Chern number on each hemisphere. 

To ensure the wavefunction is merely locally singular, we consider $|\xi^{(e,I)}({\bf k})\rangle$ for $\theta\in [\frac{\pi}{4},\frac{3\pi}{4}]$, and choose a different gauge choice for $\theta\in [0,\frac{\pi}{4}]\cup[\frac{3\pi}{4},\pi]$ with $|\xi^{(e,II)}({\bf k})\rangle = (-\cos \theta, \sin\theta e^{i\phi})^T$.
Straightforward calculations lead to:
\begin{equation}
\label{eq:deltaeff}
    \Delta_{\text{eff}}^{(I)}= d_x e^{i\phi}\sin 2\theta,\ \  \Delta_{\text{eff}}^{(II)}= d_x e^{-i\phi}\sin 2\theta.
\end{equation}
Apparently, $\Delta_{\text{eff}}$ features two zeros on ${\cal M}_+$, one at the north pole and another at the equator. Based on the form of $\Delta_{\text{eff}}^{(II)}$, it is easy to see that the vorticity of the north-pole zero is $\nu_0 = -1$. Meanwhile, the vorticity for the equator zero is also $-1$, which can be achieved by performing a {\it clockwise} loop integral of the phase of $\Delta_{\text{eff}}^{(I)}$ at $\theta=\frac{\pi}{2}-\epsilon$~\cite{supp}. Together, we find the net vorticity $\nu=-2$ for ${\cal M}_+$, which is consistent with the prediction of the Chern-vorticity theorem.     

We now make a few remarks. First of all, the vorticity-induced pairing zeros always exist, regardless of the detailed form of $d_x({\bf k})$. The zeros of $d_x({\bf k})$ itself may lead to additional zeros of $\Delta_{\text{eff}}$, beyond the topologically obstructed ones. Second, we highlight that the counting of $\nu$ for intra-FS pairing does depend on a special gauge choice with which ${\cal M}_{\pm}$ are effectively closed. Crucially, $M_z$ symmetry guarantees the existence of such a gauge in our case~\cite{supp}, ensuring the $\nu$ counting is always possible. Apparently, the nodal structure of $\Delta_{\text{eff}}$ must be gauge-invariant. Therefore, applying the Chern-vorticity theorem under a proper gauge thoroughly informs the obstructed pairing zeros that are gauge-independent.    

\begin{figure}[t]
    \centering
    \includegraphics[width=1\columnwidth]{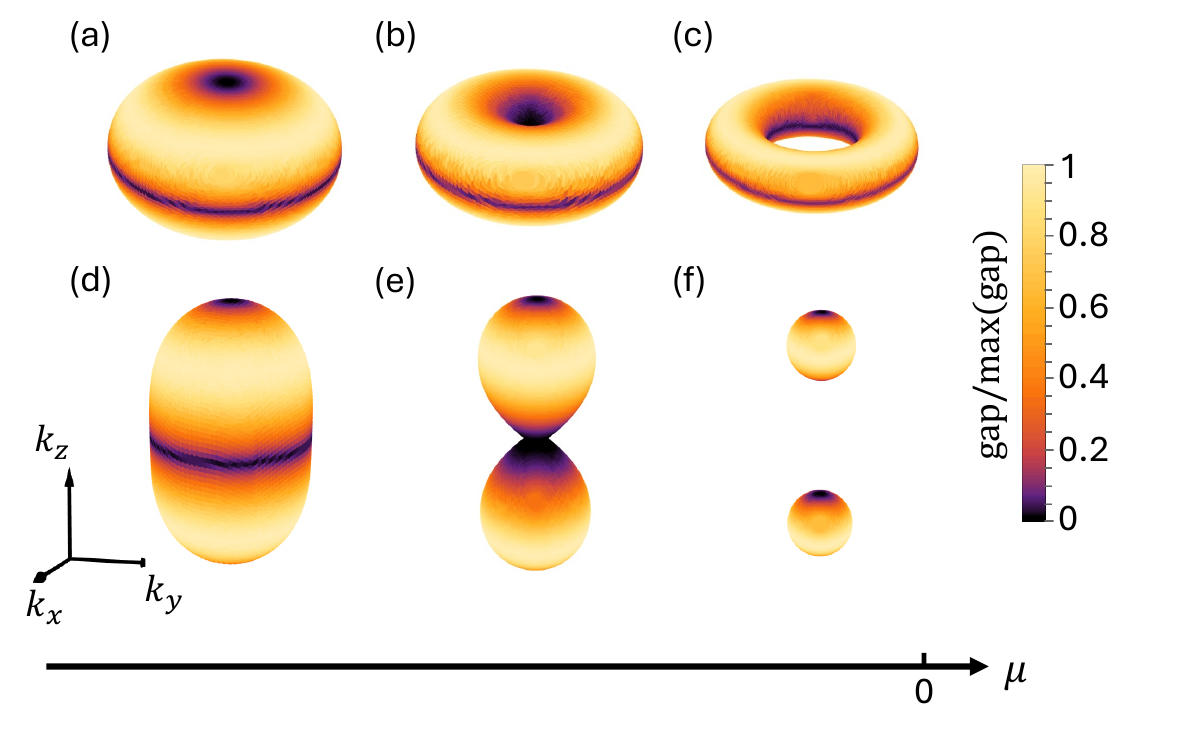}
    \caption{Superconducting gap on the FS. (a)-(c) shows the projected pairing gap for NLSM ($\Sigma=-0.1$) with $\mu=-0.2,-0.1,-0.05$, respectively. (d)-(f) shows the projected pairing gap for WSM ($\Sigma=0.1$) with $\mu=-0.2,-0.1,-0.05$, respectively. Note that a Lifshitz transition occurs at $|\mu|=|\Sigma|$ where the number of FSs changes.}
    \label{fig:FSdeform}
\end{figure}

Finally, let us take the $^2E_u$ pairing $d_x({\bf k}) = k_+$ as an example to numerically explore the nodal nature of the obstructed SC phases. In Fig.~\ref{fig:FSdeform}, we plot the pairing gap function on the corresponding FS and track its evolution as a function of $\Sigma$ and $\mu$, where the BdG Weyl nodes and nodal loops are highlighted in black. When $|\mu|>|\Sigma|$, we find a single sphere-like FS for both NLSM ($\Sigma<0$) and WSM ($\Sigma>0$), which exhibits both a BdG nodal loop at the equator and a pair of BdG Weyl nodes at the poles. This directly follows our analysis of vorticity counting. Further reducing $|\mu|$ triggers a Lifshitz transition for NLSM, which makes the point nodes merge. Surprisingly, the merging of these oppositely charged Weyl nodes leads to a second nodal loop at $k_z=0$. This is in contrast with the expected pair annihilation process in conventional Weyl SCs, as shown in the SM~\cite{supp}. Meanwhile, a similar Lifshitz transition for WSM shrinks the BdG nodal loop to a point, which further splits into a pair of Weyl nodes. This robust conversion between Weyl nodes and nodal loops for the BdG bands directly arises from the effective closedness of ${\cal M}_{\pm}$~\cite{supp}, which is reminiscent of similar behaviors in its normal state $h_0$~\cite{Sun2018conversion}.

\magenta{\it Application to $j=3/2$ SCs. |} As a real-world example, we now demonstrate that the BdG line nodes of half-Heusler-based $j=3/2$ SCs such as YPtBi and LuPdBi are deeply rooted in the mechanism of dipole obstruction. 

\begin{figure}[t]
    \centering
    \includegraphics[width=1\columnwidth]{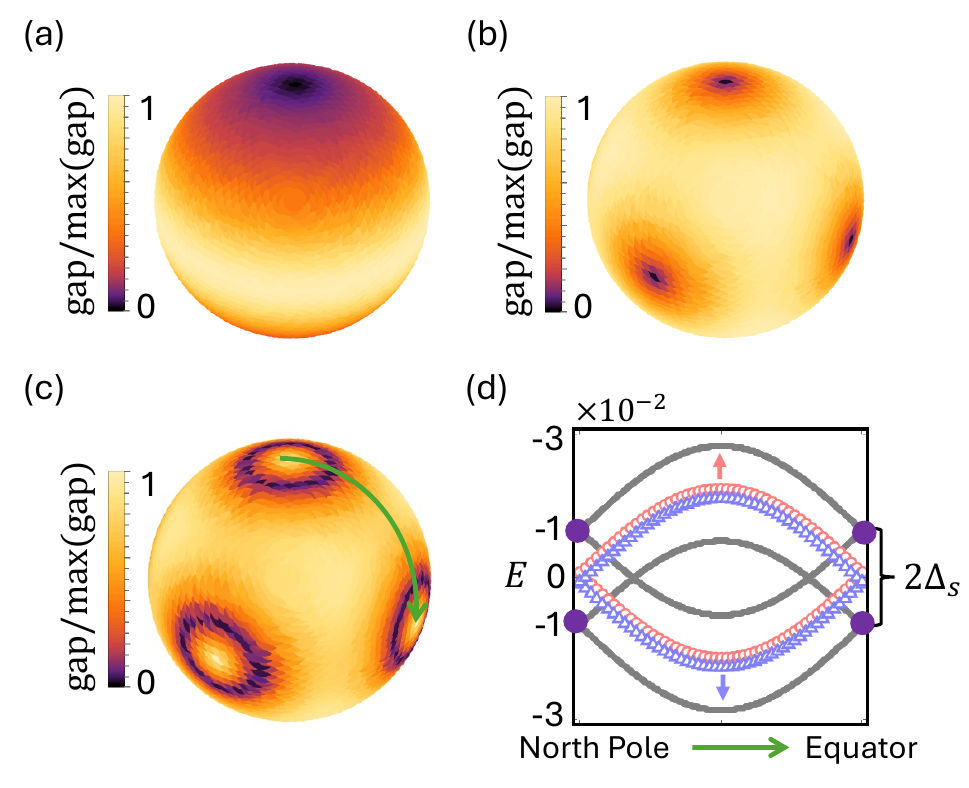}
    \caption{Superconducting gap over the FS of (a) a Dirac-dipole semimetal; (b) a Luttinger semimetal with $(\Delta_{s},\Delta_{p})=(0,0.05)$; and (c) a Luttinger semimetal with $(\Delta_{s},\Delta_{p})=(0.01,0.05)$, at $\mu=-0.2$.  (d) Spectrum along a path on FS indicated by the green arrow in (c). The purple dots highlight the band degeneracies induced by the dipole obstruction.  }
    \label{fig:DDLSM}
\end{figure}

We start with a generalized Luttinger-Kohn Hamiltonian as a normal state of interest~\cite{hu2023topological,Zhu2023delicate}, 
\begin{eqnarray}
\label{eq:generalizedLSM}
    h_{L}^{(\alpha)}({\bf k}) &=& -\sqrt{3} \alpha (k_x^2-k_y^2) \gamma_1 - 2\sqrt{3} \alpha k_x k_y \gamma_2 
    \nonumber \\
    &-&2\sqrt{3} k_z k_x \gamma_3 - 2\sqrt{3} k_z k_y\gamma_4 + M({\bf k}) \gamma_5,
\end{eqnarray} 
where $\gamma_1 = \tau_x \sigma_0$,  $\gamma_2 = \tau_y \sigma_0$, $\gamma_3 = \tau_z \sigma_x$,
$\gamma_4 = \tau_z \sigma_y$, $\gamma_5 = \tau_z \sigma_z$,
and $M({\bf k}) = k_x^2 + k_y^2 - 2k_z^2$. When $\alpha=1$, $h^{(1)}_{L}$ recovers the standard isotropic Luttinger semimetal, which nicely captures the low-energy normal state of general $j=3/2$ half-Heusler SCs. While several candidate pairing channels have been theoretically proposed~\cite{yang2017majorana,roy2019topo,boettcher2018SC,yu2018singlet,venderbos2018pairing}, a recent experiment reports a doping-dependent nodeless-to-nodal transition in LuPdBi~\cite{tuning2021ishihara}. This observation strongly supports the mixed-parity singlet-septet pairing $\tilde{\Delta}({\bf k}) = \tilde{\Delta}_s + \tilde{\Delta}_p$~\cite{pairing2016brydon}, with $\tilde{\Delta}_s = i\Delta_{s}\tau_{x}\sigma_{y}$ and
\begin{eqnarray}
\label{eq:spin3/2pairing}
\tilde{\Delta}_p = \Delta_{p} && \bigg(\frac{3}{4}k_-\tau_{z}\sigma_{+} + \frac{3}{4}k_{+}\tau_{z}\sigma_{-} + \frac{\sqrt{3}}{2}k_z\tau_{0}\sigma_{x} \nonumber \\
&&+ \frac{\sqrt{3}}{4}k_+\tau_{x}\sigma_{+}-\frac{\sqrt{3}}{4}k_-\tau_{x}\sigma_{-}\bigg),
\end{eqnarray}
which belongs to the $A_1$ irrep of $T_d$ group. Specifically, with just $\tilde{\Delta}_p$, the BdG spectrum of $h^{(1)}_{L}$ displays point nodes at $k_{x,y,z}$ axes [c.f. Fig.~\ref{fig:DDLSM}(b)]. Turning on $\tilde{\Delta}_s$ further inflates the point nodes into nodal loops, eventually leading to a full gap when $\tilde{\Delta}_s$ dominates. However, why the $p$-wave pairing can lead to point nodes can be quite puzzling. For example, let us set $\Delta_s=0$ and focus on the BdG physics along the $k_z$ axis. While $\tilde{\Delta}_p \sim k_z\tau_{0}\sigma_{x}\neq 0$ for $k_z\neq 0$, the existence of point nodes implies that the projection of $\tilde{\Delta}_p(0,0,k_z)$ onto the FS must vanish.     

To trace the origin of these point nodes, we note that when $\alpha=0$, $h_{L}^{(0)}({\bf k})\sim \tau_z\otimes h_0({\bf k})$ exactly describes a {\it Dirac-dipole semimetal}~\cite{Zhu2023spin} that comprises two decoupled copies of the Berry-dipole Hamiltonian at $\Sigma=0$, up to some parameter rescaling. The block-diagonal nature of $h_{L}^{(0)}({\bf k})$ suggests an emergent spin conservation symmetry $U(1)_s$ generated by $S_z=\tau_z \sigma_0$, with which the two Berry-dipole blocks are carrying opposite spins. Note that an $\alpha \neq 0$ generally reduces $U(1)_s$ to the time-reversal symmetry $\mathcal{T}=i\tau_{x}\sigma_{y}{\cal K}$, with ${\cal K}$ the complex conjugation. Besides, $h_{L}^{(0)}({\bf k})$ fully inherits the $C_{4h}$ symmetry of $h_0$ with an updated rotation generator $J_z=\text{diag}(\frac{3}{2},\frac{1}{2},-\frac{1}{2},-\frac{3}{2})$ and $M_z=i\tau_0\sigma_z$. For a finite $\mu$, the Dirac-dipole semimetal features a pair of decoupled FSs, one for each spin sector. Hence, all intra-FS pairings are intra-spin pairings, and vice versa.  

Notably, the first three terms of $\tilde{\Delta}_p(\mathbf{k})$ describe the intra-spin pairing process. Within each spin sector, these pairings correspond to the $B_u$ pairings in Table~\ref{tab:pairing_class_BD}, all of which are hence {\it dipole-obstructed} for $h_{L}^{(0)}({\bf k})$. Applying the Chern-vorticity theorem in Eq.~\ref{eq:Chern-vort}, we find this set of intra-spin pairing terms to carry a nontrivial pairing vorticity of $\nu=-1$~\cite{supp}, which strictly enforces pairing zeros (or equivalently point nodes) on both the north and south poles of the FS. Meanwhile, the remaining pairing terms of $\tilde{\Delta}_p(\mathbf{k})$ are inter-spin and exhibit no obvious vortex pattern. Nonetheless, both inter-spin pairings vanish on the $k_z$ axis and their existence is thus invisible to the dipole-obstructed pairing zeros. In Fig.~\ref{fig:DDLSM}(a), we numerically confirm the predicted pairing zeros of $\tilde{\Delta}_p(\mathbf{k})$ by mapping out the pairing gap on the FS of the Dirac-dipole semimetal.

We now revisit the isotropic limit ($\alpha=1$) that the $j=3/2$ SCs follow. Since all $\alpha$-relevant terms in $h_L^{(\alpha)}$ vanish on the $k_z$ axis, turning on $\alpha$ will {\it not} disrupt the obstructed pairing zeros in the Dirac-dipole limit. Moreover, $h_{L}^{(1)}({\bf k})$ respects a larger symmetry group (e.g., $T_d$) than that of $h_{L}^{(0)}({\bf k})$, which features a three-fold rotation $C_{3,111}$ that permutes $k_{x},k_{y}$, and $k_{z}$. Hence, BdG point nodes should also emerge at both $k_x$ and $k_y$ axes, which is explicitly confirmed in Fig.~\ref{fig:DDLSM}(b). In other words, $C_{3,111}$ enables us to designate any of the $k_{x,y,z}$-axes as the $C_{4}$-rotation axis for the previous Dirac-dipole discussion. In this case, {\it all of the six $p$-wave-induced point nodes for $j=3/2$ SCs have a dipole-obstructed origin.} 

Finally, let us turn on a subdominant $\tilde{\Delta}_{s}$ to achieve the BdG line nodes, as shown in Fig.~\ref{fig:DDLSM}(c). Specifically, $\tilde{\Delta}_{s}$ lifts the spin-degeneracy of FSs by breaking the inversion symmetry, which leads to accidental inter-FS band crossings that manifest as line nodes. While the dipole-obstructed pairing nodes [purple dots in Fig.~\ref{fig:DDLSM}(d)] now show up at a finite energy $E=\pm \tilde{\Delta}_{s}$, their existence ensures the BdG line nodes to necessarily show up even for an arbitrarily small $\tilde{\Delta}_{s}$. Note that we have omitted a small asymmetric spin-orbit coupling (ASOC) term in Eq.~\ref{eq:generalizedLSM}, which is intrinsic to zinc-blende materials. While the ASOC is key to mixing $s$ and $p$-wave pairings, its effect on the BdG band spectrum is similar to that of the $s$-wave pairing. A detailed discussion on the ASOC term can be found in the SM~\cite{supp}.

\magenta{\it Discussions.|} To summarize, we have established a general theoretical framework for comprehending topologically obstructed nodal pairing over a single FS. This leads us to uncover a class of Berry-dipole FSs where the SC state induced by any intra-FS pairing will be dipole-obstructed and nodal. The hidden Berry-dipole physics in the Luttinger-Kohn model further motivates us to explore the $j=3/2$ pairings in half-Heusler SCs. Focusing on the singlet-septet pairing channel, we find that the dipole obstruction contributes significantly to the BdG line nodes observed in experiments. Notably, Berry and Dirac dipoles, as well as Luttinger semimetals, have been recently established as critical points for delicate topological phases~\cite{Nelson2022delicate,Zhu2023spin,Zhu2023delicate}. Therefore, our dipole-obstructed pairing offers the first example of how delicate topological bands can decisively impact the correlated electronic orders in real-world quantum materials.  

We emphasize that our Chern-vorticity theorem goes beyond SCs and applies to general electronic orders such as charge-density waves, excitons, magnetism, etc. It would be interesting to explore possible dipole-obstructed phenomena in other non-superconducting Luttinger semimetals such as HgTe~\cite{piotrzkowski1965band} and Pr$_2$Ir$_2$O$_7$~\cite{kondo2015quadratic}. Besides, we find many inter-spin pairings for the $j=3/2$ system to feature emergent nodal structures in the BdG spectrum, despite carrying no vortex structure. As elaborated in the SM~\cite{supp} for the Dirac-dipole model, these ``unobstructed" pairing zeros arise from the geometric textures of the Bloch states on the FSs, which are beyond the scope of the Chern-vorticity theorem. How to interpret these geometry-relevant nodal pairings is an absolutely intriguing question for future research.   


\begin{acknowledgments}
We thank J. Yu, Y. Wang, L.-H. Hu, A. Alexandradinata, and Y.-M. Lu for helpful discussions. P.Z was primarily supported by the Center for Emergent Materials, an NSF MRSEC, under award number DMR-2011876. R.-X.Z. is supported by a start-up fund of the University of Tennessee. 
\end{acknowledgments}

\bibliography{refs.bib}

\begin{thebibliography}{39}%
\makeatletter
\providecommand \@ifxundefined [1]{%
 \@ifx{#1\undefined}
}%
\providecommand \@ifnum [1]{%
 \ifnum #1\expandafter \@firstoftwo
 \else \expandafter \@secondoftwo
 \fi
}%
\providecommand \@ifx [1]{%
 \ifx #1\expandafter \@firstoftwo
 \else \expandafter \@secondoftwo
 \fi
}%
\providecommand \natexlab [1]{#1}%
\providecommand \enquote  [1]{``#1''}%
\providecommand \bibnamefont  [1]{#1}%
\providecommand \bibfnamefont [1]{#1}%
\providecommand \citenamefont [1]{#1}%
\providecommand \href@noop [0]{\@secondoftwo}%
\providecommand \href [0]{\begingroup \@sanitize@url \@href}%
\providecommand \@href[1]{\@@startlink{#1}\@@href}%
\providecommand \@@href[1]{\endgroup#1\@@endlink}%
\providecommand \@sanitize@url [0]{\catcode `\\12\catcode `\$12\catcode `\&12\catcode `\#12\catcode `\^12\catcode `\_12\catcode `\%12\relax}%
\providecommand \@@startlink[1]{}%
\providecommand \@@endlink[0]{}%
\providecommand \url  [0]{\begingroup\@sanitize@url \@url }%
\providecommand \@url [1]{\endgroup\@href {#1}{\urlprefix }}%
\providecommand \urlprefix  [0]{URL }%
\providecommand \Eprint [0]{\href }%
\providecommand \doibase [0]{https://doi.org/}%
\providecommand \selectlanguage [0]{\@gobble}%
\providecommand \bibinfo  [0]{\@secondoftwo}%
\providecommand \bibfield  [0]{\@secondoftwo}%
\providecommand \translation [1]{[#1]}%
\providecommand \BibitemOpen [0]{}%
\providecommand \bibitemStop [0]{}%
\providecommand \bibitemNoStop [0]{.\EOS\space}%
\providecommand \EOS [0]{\spacefactor3000\relax}%
\providecommand \BibitemShut  [1]{\csname bibitem#1\endcsname}%
\let\auto@bib@innerbib\@empty
\bibitem [{\citenamefont {Bardeen}\ \emph {et~al.}(1957)\citenamefont {Bardeen}, \citenamefont {Cooper},\ and\ \citenamefont {Schrieffer}}]{BCS1957}%
  \BibitemOpen
  \bibfield  {author} {\bibinfo {author} {\bibfnamefont {J.}~\bibnamefont {Bardeen}}, \bibinfo {author} {\bibfnamefont {L.~N.}\ \bibnamefont {Cooper}},\ and\ \bibinfo {author} {\bibfnamefont {J.~R.}\ \bibnamefont {Schrieffer}},\ }\bibfield  {title} {\bibinfo {title} {Microscopic theory of superconductivity},\ }\href {https://doi.org/10.1103/PhysRev.106.162} {\bibfield  {journal} {\bibinfo  {journal} {Phys. Rev.}\ }\textbf {\bibinfo {volume} {106}},\ \bibinfo {pages} {162} (\bibinfo {year} {1957})}\BibitemShut {NoStop}%
\bibitem [{\citenamefont {Sigrist}\ and\ \citenamefont {Ueda}(1991)}]{sigrist1991RMP}%
  \BibitemOpen
  \bibfield  {author} {\bibinfo {author} {\bibfnamefont {M.}~\bibnamefont {Sigrist}}\ and\ \bibinfo {author} {\bibfnamefont {K.}~\bibnamefont {Ueda}},\ }\bibfield  {title} {\bibinfo {title} {Phenomenological theory of unconventional superconductivity},\ }\href {https://doi.org/10.1103/RevModPhys.63.239} {\bibfield  {journal} {\bibinfo  {journal} {Rev. Mod. Phys.}\ }\textbf {\bibinfo {volume} {63}},\ \bibinfo {pages} {239} (\bibinfo {year} {1991})}\BibitemShut {NoStop}%
\bibitem [{\citenamefont {Tsuei}\ and\ \citenamefont {Kirtley}(2000)}]{tsuei2000RMP}%
  \BibitemOpen
  \bibfield  {author} {\bibinfo {author} {\bibfnamefont {C.~C.}\ \bibnamefont {Tsuei}}\ and\ \bibinfo {author} {\bibfnamefont {J.~R.}\ \bibnamefont {Kirtley}},\ }\bibfield  {title} {\bibinfo {title} {Pairing symmetry in cuprate superconductors},\ }\href {https://doi.org/10.1103/RevModPhys.72.969} {\bibfield  {journal} {\bibinfo  {journal} {Rev. Mod. Phys.}\ }\textbf {\bibinfo {volume} {72}},\ \bibinfo {pages} {969} (\bibinfo {year} {2000})}\BibitemShut {NoStop}%
\bibitem [{\citenamefont {Kim}\ \emph {et~al.}(2018)\citenamefont {Kim}, \citenamefont {Wang}, \citenamefont {Nakajima}, \citenamefont {Hu}, \citenamefont {Ziemak}, \citenamefont {Syers}, \citenamefont {Wang}, \citenamefont {Hodovanets}, \citenamefont {Denlinger}, \citenamefont {Brydon} \emph {et~al.}}]{kim2018beyond}%
  \BibitemOpen
  \bibfield  {author} {\bibinfo {author} {\bibfnamefont {H.}~\bibnamefont {Kim}}, \bibinfo {author} {\bibfnamefont {K.}~\bibnamefont {Wang}}, \bibinfo {author} {\bibfnamefont {Y.}~\bibnamefont {Nakajima}}, \bibinfo {author} {\bibfnamefont {R.}~\bibnamefont {Hu}}, \bibinfo {author} {\bibfnamefont {S.}~\bibnamefont {Ziemak}}, \bibinfo {author} {\bibfnamefont {P.}~\bibnamefont {Syers}}, \bibinfo {author} {\bibfnamefont {L.}~\bibnamefont {Wang}}, \bibinfo {author} {\bibfnamefont {H.}~\bibnamefont {Hodovanets}}, \bibinfo {author} {\bibfnamefont {J.~D.}\ \bibnamefont {Denlinger}}, \bibinfo {author} {\bibfnamefont {P.~M.}\ \bibnamefont {Brydon}}, \emph {et~al.},\ }\bibfield  {title} {\bibinfo {title} {Beyond triplet: Unconventional superconductivity in a spin-3/2 topological semimetal},\ }\href {https://www.science.org/doi/10.1126/sciadv.aao4513} {\bibfield  {journal} {\bibinfo  {journal} {Science advances}\ }\textbf {\bibinfo {volume} {4}},\ \bibinfo {pages} {eaao4513} (\bibinfo {year} {2018})}\BibitemShut
  {NoStop}%
\bibitem [{\citenamefont {Damascelli}\ \emph {et~al.}(2003)\citenamefont {Damascelli}, \citenamefont {Hussain},\ and\ \citenamefont {Shen}}]{damascelli2003ARPES}%
  \BibitemOpen
  \bibfield  {author} {\bibinfo {author} {\bibfnamefont {A.}~\bibnamefont {Damascelli}}, \bibinfo {author} {\bibfnamefont {Z.}~\bibnamefont {Hussain}},\ and\ \bibinfo {author} {\bibfnamefont {Z.-X.}\ \bibnamefont {Shen}},\ }\bibfield  {title} {\bibinfo {title} {Angle-resolved photoemission studies of the cuprate superconductors},\ }\href {https://doi.org/10.1103/RevModPhys.75.473} {\bibfield  {journal} {\bibinfo  {journal} {Rev. Mod. Phys.}\ }\textbf {\bibinfo {volume} {75}},\ \bibinfo {pages} {473} (\bibinfo {year} {2003})}\BibitemShut {NoStop}%
\bibitem [{\citenamefont {Fischer}\ \emph {et~al.}(2007)\citenamefont {Fischer}, \citenamefont {Kugler}, \citenamefont {Maggio-Aprile}, \citenamefont {Berthod},\ and\ \citenamefont {Renner}}]{fischer2007STMRMP}%
  \BibitemOpen
  \bibfield  {author} {\bibinfo {author} {\bibfnamefont {O.}~\bibnamefont {Fischer}}, \bibinfo {author} {\bibfnamefont {M.}~\bibnamefont {Kugler}}, \bibinfo {author} {\bibfnamefont {I.}~\bibnamefont {Maggio-Aprile}}, \bibinfo {author} {\bibfnamefont {C.}~\bibnamefont {Berthod}},\ and\ \bibinfo {author} {\bibfnamefont {C.}~\bibnamefont {Renner}},\ }\bibfield  {title} {\bibinfo {title} {Scanning tunneling spectroscopy of high-temperature superconductors},\ }\href {https://doi.org/10.1103/RevModPhys.79.353} {\bibfield  {journal} {\bibinfo  {journal} {Rev. Mod. Phys.}\ }\textbf {\bibinfo {volume} {79}},\ \bibinfo {pages} {353} (\bibinfo {year} {2007})}\BibitemShut {NoStop}%
\bibitem [{\citenamefont {Prozorov}\ and\ \citenamefont {Giannetta}(2006)}]{prozorov2006penetration}%
  \BibitemOpen
  \bibfield  {author} {\bibinfo {author} {\bibfnamefont {R.}~\bibnamefont {Prozorov}}\ and\ \bibinfo {author} {\bibfnamefont {R.~W.}\ \bibnamefont {Giannetta}},\ }\bibfield  {title} {\bibinfo {title} {Magnetic penetration depth in unconventional superconductors},\ }\href {https://doi.org/10.1088/0953-2048/19/8/R01} {\bibfield  {journal} {\bibinfo  {journal} {Superconductor Science and Technology}\ }\textbf {\bibinfo {volume} {19}},\ \bibinfo {pages} {R41} (\bibinfo {year} {2006})}\BibitemShut {NoStop}%
\bibitem [{\citenamefont {Murakami}\ and\ \citenamefont {Nagaosa}(2003)}]{murakami2003berry}%
  \BibitemOpen
  \bibfield  {author} {\bibinfo {author} {\bibfnamefont {S.}~\bibnamefont {Murakami}}\ and\ \bibinfo {author} {\bibfnamefont {N.}~\bibnamefont {Nagaosa}},\ }\bibfield  {title} {\bibinfo {title} {Berry phase in magnetic superconductors},\ }\href {https://doi.org/10.1103/PhysRevLett.90.057002} {\bibfield  {journal} {\bibinfo  {journal} {Phys. Rev. Lett.}\ }\textbf {\bibinfo {volume} {90}},\ \bibinfo {pages} {057002} (\bibinfo {year} {2003})}\BibitemShut {NoStop}%
\bibitem [{\citenamefont {Li}\ and\ \citenamefont {Haldane}(2018)}]{Li2018topological}%
  \BibitemOpen
  \bibfield  {author} {\bibinfo {author} {\bibfnamefont {Y.}~\bibnamefont {Li}}\ and\ \bibinfo {author} {\bibfnamefont {F.~D.~M.}\ \bibnamefont {Haldane}},\ }\bibfield  {title} {\bibinfo {title} {Topological nodal cooper pairing in doped weyl metals},\ }\href {https://doi.org/10.1103/PhysRevLett.120.067003} {\bibfield  {journal} {\bibinfo  {journal} {Phys. Rev. Lett.}\ }\textbf {\bibinfo {volume} {120}},\ \bibinfo {pages} {067003} (\bibinfo {year} {2018})}\BibitemShut {NoStop}%
\bibitem [{\citenamefont {Wang}\ and\ \citenamefont {Ye}(2016)}]{wang2016densitywave}%
  \BibitemOpen
  \bibfield  {author} {\bibinfo {author} {\bibfnamefont {Y.}~\bibnamefont {Wang}}\ and\ \bibinfo {author} {\bibfnamefont {P.}~\bibnamefont {Ye}},\ }\bibfield  {title} {\bibinfo {title} {Topological density-wave states in a particle-hole symmetric weyl metal},\ }\href {https://doi.org/10.1103/PhysRevB.94.075115} {\bibfield  {journal} {\bibinfo  {journal} {Phys. Rev. B}\ }\textbf {\bibinfo {volume} {94}},\ \bibinfo {pages} {075115} (\bibinfo {year} {2016})}\BibitemShut {NoStop}%
\bibitem [{\citenamefont {Mu\~noz}\ \emph {et~al.}(2020)\citenamefont {Mu\~noz}, \citenamefont {Soto-Garrido},\ and\ \citenamefont {Juri\ifmmode \check{c}\else \v{c}\fi{}i\ifmmode~\acute{c}\else \'{c}\fi{}}}]{Munoz2020monopole}%
  \BibitemOpen
  \bibfield  {author} {\bibinfo {author} {\bibfnamefont {E.}~\bibnamefont {Mu\~noz}}, \bibinfo {author} {\bibfnamefont {R.}~\bibnamefont {Soto-Garrido}},\ and\ \bibinfo {author} {\bibfnamefont {V.}~\bibnamefont {Juri\ifmmode \check{c}\else \v{c}\fi{}i\ifmmode~\acute{c}\else \'{c}\fi{}}},\ }\bibfield  {title} {\bibinfo {title} {Monopole versus spherical harmonic superconductors: Topological repulsion, coexistence, and stability},\ }\href {https://doi.org/10.1103/PhysRevB.102.195121} {\bibfield  {journal} {\bibinfo  {journal} {Phys. Rev. B}\ }\textbf {\bibinfo {volume} {102}},\ \bibinfo {pages} {195121} (\bibinfo {year} {2020})}\BibitemShut {NoStop}%
\bibitem [{\citenamefont {Yu}\ \emph {et~al.}(2022)\citenamefont {Yu}, \citenamefont {Chen},\ and\ \citenamefont {Das~Sarma}}]{Yu2022euler}%
  \BibitemOpen
  \bibfield  {author} {\bibinfo {author} {\bibfnamefont {J.}~\bibnamefont {Yu}}, \bibinfo {author} {\bibfnamefont {Y.-A.}\ \bibnamefont {Chen}},\ and\ \bibinfo {author} {\bibfnamefont {S.}~\bibnamefont {Das~Sarma}},\ }\bibfield  {title} {\bibinfo {title} {Euler-obstructed cooper pairing: Nodal superconductivity and hinge majorana zero modes},\ }\href {https://doi.org/10.1103/PhysRevB.105.104515} {\bibfield  {journal} {\bibinfo  {journal} {Phys. Rev. B}\ }\textbf {\bibinfo {volume} {105}},\ \bibinfo {pages} {104515} (\bibinfo {year} {2022})}\BibitemShut {NoStop}%
\bibitem [{\citenamefont {Yu}\ \emph {et~al.}(2023)\citenamefont {Yu}, \citenamefont {Xie}, \citenamefont {Wu},\ and\ \citenamefont {Das~Sarma}}]{yu2023euler}%
  \BibitemOpen
  \bibfield  {author} {\bibinfo {author} {\bibfnamefont {J.}~\bibnamefont {Yu}}, \bibinfo {author} {\bibfnamefont {M.}~\bibnamefont {Xie}}, \bibinfo {author} {\bibfnamefont {F.}~\bibnamefont {Wu}},\ and\ \bibinfo {author} {\bibfnamefont {S.}~\bibnamefont {Das~Sarma}},\ }\bibfield  {title} {\bibinfo {title} {Euler-obstructed nematic nodal superconductivity in twisted bilayer graphene},\ }\href {https://doi.org/10.1103/PhysRevB.107.L201106} {\bibfield  {journal} {\bibinfo  {journal} {Phys. Rev. B}\ }\textbf {\bibinfo {volume} {107}},\ \bibinfo {pages} {L201106} (\bibinfo {year} {2023})}\BibitemShut {NoStop}%
\bibitem [{\citenamefont {Wang}\ \emph {et~al.}(2024)\citenamefont {Wang}, \citenamefont {Zhou}, \citenamefont {Peng}, \citenamefont {Lian},\ and\ \citenamefont {Song}}]{wang2024molecular}%
  \BibitemOpen
  \bibfield  {author} {\bibinfo {author} {\bibfnamefont {Y.-J.}\ \bibnamefont {Wang}}, \bibinfo {author} {\bibfnamefont {G.-D.}\ \bibnamefont {Zhou}}, \bibinfo {author} {\bibfnamefont {S.-Y.}\ \bibnamefont {Peng}}, \bibinfo {author} {\bibfnamefont {B.}~\bibnamefont {Lian}},\ and\ \bibinfo {author} {\bibfnamefont {Z.-D.}\ \bibnamefont {Song}},\ }\bibfield  {title} {\bibinfo {title} {Molecular pairing in twisted bilayer graphene superconductivity},\ }\href {https://doi.org/10.48550/arXiv.2402.00869} {\bibfield  {journal} {\bibinfo  {journal} {arXiv preprint arXiv:2402.00869}\ } (\bibinfo {year} {2024})}\BibitemShut {NoStop}%
\bibitem [{\citenamefont {Sun}\ and\ \citenamefont {Li}(2020)}]{Sun2020mathbb}%
  \BibitemOpen
  \bibfield  {author} {\bibinfo {author} {\bibfnamefont {C.}~\bibnamefont {Sun}}\ and\ \bibinfo {author} {\bibfnamefont {Y.}~\bibnamefont {Li}},\ }\bibfield  {title} {\bibinfo {title} {$\mathbb{Z}_2$ topologically obstructed superconducting order},\ }\href {https://doi.org/10.48550/arXiv.2009.07263} {\bibfield  {journal} {\bibinfo  {journal} {arXiv preprint arXiv:2009.07263}\ } (\bibinfo {year} {2020})}\BibitemShut {NoStop}%
\bibitem [{\citenamefont {Nelson}\ \emph {et~al.}(2022)\citenamefont {Nelson}, \citenamefont {Neupert}, \citenamefont {Alexandradinata},\ and\ \citenamefont {Bzdu\ifmmode~\check{s}\else \v{s}\fi{}ek}}]{Nelson2022delicate}%
  \BibitemOpen
  \bibfield  {author} {\bibinfo {author} {\bibfnamefont {A.}~\bibnamefont {Nelson}}, \bibinfo {author} {\bibfnamefont {T.}~\bibnamefont {Neupert}}, \bibinfo {author} {\bibfnamefont {A.}~\bibnamefont {Alexandradinata}},\ and\ \bibinfo {author} {\bibfnamefont {T.~c.~v.}\ \bibnamefont {Bzdu\ifmmode~\check{s}\else \v{s}\fi{}ek}},\ }\bibfield  {title} {\bibinfo {title} {Delicate topology protected by rotation symmetry: Crystalline hopf insulators and beyond},\ }\href {https://doi.org/10.1103/PhysRevB.106.075124} {\bibfield  {journal} {\bibinfo  {journal} {Phys. Rev. B}\ }\textbf {\bibinfo {volume} {106}},\ \bibinfo {pages} {075124} (\bibinfo {year} {2022})}\BibitemShut {NoStop}%
\bibitem [{\citenamefont {Butch}\ \emph {et~al.}(2011)\citenamefont {Butch}, \citenamefont {Syers}, \citenamefont {Kirshenbaum}, \citenamefont {Hope},\ and\ \citenamefont {Paglione}}]{butch2011SC}%
  \BibitemOpen
  \bibfield  {author} {\bibinfo {author} {\bibfnamefont {N.~P.}\ \bibnamefont {Butch}}, \bibinfo {author} {\bibfnamefont {P.}~\bibnamefont {Syers}}, \bibinfo {author} {\bibfnamefont {K.}~\bibnamefont {Kirshenbaum}}, \bibinfo {author} {\bibfnamefont {A.~P.}\ \bibnamefont {Hope}},\ and\ \bibinfo {author} {\bibfnamefont {J.}~\bibnamefont {Paglione}},\ }\bibfield  {title} {\bibinfo {title} {Superconductivity in the topological semimetal yptbi},\ }\href {https://doi.org/10.1103/PhysRevB.84.220504} {\bibfield  {journal} {\bibinfo  {journal} {Phys. Rev. B}\ }\textbf {\bibinfo {volume} {84}},\ \bibinfo {pages} {220504} (\bibinfo {year} {2011})}\BibitemShut {NoStop}%
\bibitem [{\citenamefont {Nakajima}\ \emph {et~al.}(2015)\citenamefont {Nakajima}, \citenamefont {Hu}, \citenamefont {Kirshenbaum}, \citenamefont {Hughes}, \citenamefont {Syers}, \citenamefont {Wang}, \citenamefont {Wang}, \citenamefont {Wang}, \citenamefont {Saha}, \citenamefont {Pratt} \emph {et~al.}}]{nakajima2015topological}%
  \BibitemOpen
  \bibfield  {author} {\bibinfo {author} {\bibfnamefont {Y.}~\bibnamefont {Nakajima}}, \bibinfo {author} {\bibfnamefont {R.}~\bibnamefont {Hu}}, \bibinfo {author} {\bibfnamefont {K.}~\bibnamefont {Kirshenbaum}}, \bibinfo {author} {\bibfnamefont {A.}~\bibnamefont {Hughes}}, \bibinfo {author} {\bibfnamefont {P.}~\bibnamefont {Syers}}, \bibinfo {author} {\bibfnamefont {X.}~\bibnamefont {Wang}}, \bibinfo {author} {\bibfnamefont {K.}~\bibnamefont {Wang}}, \bibinfo {author} {\bibfnamefont {R.}~\bibnamefont {Wang}}, \bibinfo {author} {\bibfnamefont {S.~R.}\ \bibnamefont {Saha}}, \bibinfo {author} {\bibfnamefont {D.}~\bibnamefont {Pratt}}, \emph {et~al.},\ }\bibfield  {title} {\bibinfo {title} {Topological r pdbi half-heusler semimetals: A new family of noncentrosymmetric magnetic superconductors},\ }\href {https://www.science.org/doi/10.1126/sciadv.1500242} {\bibfield  {journal} {\bibinfo  {journal} {Science advances}\ }\textbf {\bibinfo {volume} {1}},\ \bibinfo {pages} {e1500242} (\bibinfo {year}
  {2015})}\BibitemShut {NoStop}%
\bibitem [{\citenamefont {Pavlosiuk}\ \emph {et~al.}(2015)\citenamefont {Pavlosiuk}, \citenamefont {Kaczorowski},\ and\ \citenamefont {Wi{\'s}niewski}}]{pavlosiuk2015shubnikov}%
  \BibitemOpen
  \bibfield  {author} {\bibinfo {author} {\bibfnamefont {O.}~\bibnamefont {Pavlosiuk}}, \bibinfo {author} {\bibfnamefont {D.}~\bibnamefont {Kaczorowski}},\ and\ \bibinfo {author} {\bibfnamefont {P.}~\bibnamefont {Wi{\'s}niewski}},\ }\bibfield  {title} {\bibinfo {title} {Shubnikov-de haas oscillations, weak antilocalization effect and large linear magnetoresistance in the putative topological superconductor lupdbi},\ }\href {https://doi.org/10.1038/srep09158} {\bibfield  {journal} {\bibinfo  {journal} {Scientific reports}\ }\textbf {\bibinfo {volume} {5}},\ \bibinfo {pages} {9158} (\bibinfo {year} {2015})}\BibitemShut {NoStop}%
\bibitem [{\citenamefont {Ishihara}\ \emph {et~al.}(2021)\citenamefont {Ishihara}, \citenamefont {Takenaka}, \citenamefont {Miao}, \citenamefont {Mizukami}, \citenamefont {Hashimoto}, \citenamefont {Yamashita}, \citenamefont {Konczykowski}, \citenamefont {Masuki}, \citenamefont {Hirayama}, \citenamefont {Nomoto}, \citenamefont {Arita}, \citenamefont {Pavlosiuk}, \citenamefont {Wi\ifmmode~\acute{s}\else \'{s}\fi{}niewski}, \citenamefont {Kaczorowski},\ and\ \citenamefont {Shibauchi}}]{tuning2021ishihara}%
  \BibitemOpen
  \bibfield  {author} {\bibinfo {author} {\bibfnamefont {K.}~\bibnamefont {Ishihara}}, \bibinfo {author} {\bibfnamefont {T.}~\bibnamefont {Takenaka}}, \bibinfo {author} {\bibfnamefont {Y.}~\bibnamefont {Miao}}, \bibinfo {author} {\bibfnamefont {Y.}~\bibnamefont {Mizukami}}, \bibinfo {author} {\bibfnamefont {K.}~\bibnamefont {Hashimoto}}, \bibinfo {author} {\bibfnamefont {M.}~\bibnamefont {Yamashita}}, \bibinfo {author} {\bibfnamefont {M.}~\bibnamefont {Konczykowski}}, \bibinfo {author} {\bibfnamefont {R.}~\bibnamefont {Masuki}}, \bibinfo {author} {\bibfnamefont {M.}~\bibnamefont {Hirayama}}, \bibinfo {author} {\bibfnamefont {T.}~\bibnamefont {Nomoto}}, \bibinfo {author} {\bibfnamefont {R.}~\bibnamefont {Arita}}, \bibinfo {author} {\bibfnamefont {O.}~\bibnamefont {Pavlosiuk}}, \bibinfo {author} {\bibfnamefont {P.}~\bibnamefont {Wi\ifmmode~\acute{s}\else \'{s}\fi{}niewski}}, \bibinfo {author} {\bibfnamefont {D.}~\bibnamefont {Kaczorowski}},\ and\ \bibinfo {author} {\bibfnamefont {T.}~\bibnamefont {Shibauchi}},\
  }\bibfield  {title} {\bibinfo {title} {Tuning the parity mixing of singlet-septet pairing in a half-heusler superconductor},\ }\href {https://doi.org/10.1103/PhysRevX.11.041048} {\bibfield  {journal} {\bibinfo  {journal} {Phys. Rev. X}\ }\textbf {\bibinfo {volume} {11}},\ \bibinfo {pages} {041048} (\bibinfo {year} {2021})}\BibitemShut {NoStop}%
\bibitem [{\citenamefont {Brydon}\ \emph {et~al.}(2016)\citenamefont {Brydon}, \citenamefont {Wang}, \citenamefont {Weinert},\ and\ \citenamefont {Agterberg}}]{pairing2016brydon}%
  \BibitemOpen
  \bibfield  {author} {\bibinfo {author} {\bibfnamefont {P.~M.~R.}\ \bibnamefont {Brydon}}, \bibinfo {author} {\bibfnamefont {L.}~\bibnamefont {Wang}}, \bibinfo {author} {\bibfnamefont {M.}~\bibnamefont {Weinert}},\ and\ \bibinfo {author} {\bibfnamefont {D.~F.}\ \bibnamefont {Agterberg}},\ }\bibfield  {title} {\bibinfo {title} {Pairing of $j=3/2$ fermions in half-heusler superconductors},\ }\href {https://doi.org/10.1103/PhysRevLett.116.177001} {\bibfield  {journal} {\bibinfo  {journal} {Phys. Rev. Lett.}\ }\textbf {\bibinfo {volume} {116}},\ \bibinfo {pages} {177001} (\bibinfo {year} {2016})}\BibitemShut {NoStop}%
\bibitem [{sup()}]{supp}%
  \BibitemOpen
  \bibinfo {note} {See the Supplemental Material for (i) a detailed proof of the Chern-vorticity theorem; (ii) detailed analysis for general pairings in Berry-dipole semimetals and inter-spin pairings in Dirac-dipole semimetals; and (iii) case studies on WSM, Berry dipole SC, and perturbations from ASOC.}\BibitemShut {Stop}%
\bibitem [{\citenamefont {Bobrow}\ \emph {et~al.}(2020)\citenamefont {Bobrow}, \citenamefont {Sun},\ and\ \citenamefont {Li}}]{bobrow2020monopole}%
  \BibitemOpen
  \bibfield  {author} {\bibinfo {author} {\bibfnamefont {E.}~\bibnamefont {Bobrow}}, \bibinfo {author} {\bibfnamefont {C.}~\bibnamefont {Sun}},\ and\ \bibinfo {author} {\bibfnamefont {Y.}~\bibnamefont {Li}},\ }\bibfield  {title} {\bibinfo {title} {Monopole charge density wave states in weyl semimetals},\ }\href {https://doi.org/10.1103/PhysRevResearch.2.012078} {\bibfield  {journal} {\bibinfo  {journal} {Phys. Rev. Res.}\ }\textbf {\bibinfo {volume} {2}},\ \bibinfo {pages} {012078} (\bibinfo {year} {2020})}\BibitemShut {NoStop}%
\bibitem [{\citenamefont {Bultinck}\ \emph {et~al.}(2020)\citenamefont {Bultinck}, \citenamefont {Chatterjee},\ and\ \citenamefont {Zaletel}}]{bultinck2020mechanism}%
  \BibitemOpen
  \bibfield  {author} {\bibinfo {author} {\bibfnamefont {N.}~\bibnamefont {Bultinck}}, \bibinfo {author} {\bibfnamefont {S.}~\bibnamefont {Chatterjee}},\ and\ \bibinfo {author} {\bibfnamefont {M.~P.}\ \bibnamefont {Zaletel}},\ }\bibfield  {title} {\bibinfo {title} {Mechanism for anomalous hall ferromagnetism in twisted bilayer graphene},\ }\href {https://doi.org/10.1103/PhysRevLett.124.166601} {\bibfield  {journal} {\bibinfo  {journal} {Phys. Rev. Lett.}\ }\textbf {\bibinfo {volume} {124}},\ \bibinfo {pages} {166601} (\bibinfo {year} {2020})}\BibitemShut {NoStop}%
\bibitem [{\citenamefont {Zhu}\ and\ \citenamefont {Alexandradinata}(2023)}]{zhu2023anomalous}%
  \BibitemOpen
  \bibfield  {author} {\bibinfo {author} {\bibfnamefont {P.}~\bibnamefont {Zhu}}\ and\ \bibinfo {author} {\bibfnamefont {A.}~\bibnamefont {Alexandradinata}},\ }\bibfield  {title} {\bibinfo {title} {Anomalous shift and optical vorticity in the steady photovoltaic current},\ }\href {https://doi.org/10.48550/arXiv.2308.08596} {\bibfield  {journal} {\bibinfo  {journal} {arXiv preprint arXiv:2308.08596}\ } (\bibinfo {year} {2023})}\BibitemShut {NoStop}%
\bibitem [{\citenamefont {Sun}\ \emph {et~al.}(2018)\citenamefont {Sun}, \citenamefont {Zhang},\ and\ \citenamefont {Bzdu\ifmmode~\check{s}\else \v{s}\fi{}ek}}]{Sun2018conversion}%
  \BibitemOpen
  \bibfield  {author} {\bibinfo {author} {\bibfnamefont {X.-Q.}\ \bibnamefont {Sun}}, \bibinfo {author} {\bibfnamefont {S.-C.}\ \bibnamefont {Zhang}},\ and\ \bibinfo {author} {\bibfnamefont {T.~c.~v.}\ \bibnamefont {Bzdu\ifmmode~\check{s}\else \v{s}\fi{}ek}},\ }\bibfield  {title} {\bibinfo {title} {Conversion rules for weyl points and nodal lines in topological media},\ }\href {https://doi.org/10.1103/PhysRevLett.121.106402} {\bibfield  {journal} {\bibinfo  {journal} {Phys. Rev. Lett.}\ }\textbf {\bibinfo {volume} {121}},\ \bibinfo {pages} {106402} (\bibinfo {year} {2018})}\BibitemShut {NoStop}%
\bibitem [{\citenamefont {Graf}\ and\ \citenamefont {Pi\'echon}(2023)}]{graf2023massless}%
  \BibitemOpen
  \bibfield  {author} {\bibinfo {author} {\bibfnamefont {A.}~\bibnamefont {Graf}}\ and\ \bibinfo {author} {\bibfnamefont {F.}~\bibnamefont {Pi\'echon}},\ }\bibfield  {title} {\bibinfo {title} {Massless multifold hopf semimetals},\ }\href {https://doi.org/10.1103/PhysRevB.108.115105} {\bibfield  {journal} {\bibinfo  {journal} {Phys. Rev. B}\ }\textbf {\bibinfo {volume} {108}},\ \bibinfo {pages} {115105} (\bibinfo {year} {2023})}\BibitemShut {NoStop}%
\bibitem [{\citenamefont {Zhuang}\ \emph {et~al.}(2024)\citenamefont {Zhuang}, \citenamefont {Zhang}, \citenamefont {Wang},\ and\ \citenamefont {Yan}}]{zhuang2024berry}%
  \BibitemOpen
  \bibfield  {author} {\bibinfo {author} {\bibfnamefont {Z.-Y.}\ \bibnamefont {Zhuang}}, \bibinfo {author} {\bibfnamefont {C.}~\bibnamefont {Zhang}}, \bibinfo {author} {\bibfnamefont {X.-J.}\ \bibnamefont {Wang}},\ and\ \bibinfo {author} {\bibfnamefont {Z.}~\bibnamefont {Yan}},\ }\bibfield  {title} {\bibinfo {title} {Berry-dipole semimetals},\ }\href {https://arxiv.org/abs/2404.10049} {\bibfield  {journal} {\bibinfo  {journal} {arXiv preprint arXiv:2404.10049}\ } (\bibinfo {year} {2024})}\BibitemShut {NoStop}%
\bibitem [{\citenamefont {Tyner}\ and\ \citenamefont {Sur}(2024)}]{tyner2024dipolar}%
  \BibitemOpen
  \bibfield  {author} {\bibinfo {author} {\bibfnamefont {A.~C.}\ \bibnamefont {Tyner}}\ and\ \bibinfo {author} {\bibfnamefont {S.}~\bibnamefont {Sur}},\ }\bibfield  {title} {\bibinfo {title} {Dipolar weyl semimetals},\ }\href {https://doi.org/10.1103/PhysRevB.109.L081101} {\bibfield  {journal} {\bibinfo  {journal} {Phys. Rev. B}\ }\textbf {\bibinfo {volume} {109}},\ \bibinfo {pages} {L081101} (\bibinfo {year} {2024})}\BibitemShut {NoStop}%
\bibitem [{\citenamefont {Hu}\ and\ \citenamefont {Zhang}(2023)}]{hu2023topological}%
  \BibitemOpen
  \bibfield  {author} {\bibinfo {author} {\bibfnamefont {L.-H.}\ \bibnamefont {Hu}}\ and\ \bibinfo {author} {\bibfnamefont {R.-X.}\ \bibnamefont {Zhang}},\ }\bibfield  {title} {\bibinfo {title} {Topological superconducting vortex from trivial electronic bands},\ }\href {https://doi.org/10.1038/s41467-023-36347-w} {\bibfield  {journal} {\bibinfo  {journal} {Nature Communications}\ }\textbf {\bibinfo {volume} {14}},\ \bibinfo {pages} {640} (\bibinfo {year} {2023})}\BibitemShut {NoStop}%
\bibitem [{\citenamefont {Zhu}\ and\ \citenamefont {Zhang}(2023)}]{Zhu2023delicate}%
  \BibitemOpen
  \bibfield  {author} {\bibinfo {author} {\bibfnamefont {P.}~\bibnamefont {Zhu}}\ and\ \bibinfo {author} {\bibfnamefont {R.-X.}\ \bibnamefont {Zhang}},\ }\bibfield  {title} {\bibinfo {title} {Delicate topology of luttinger semimetal},\ }\href {https://doi.org/10.48550/arXiv.2308.05793} {\bibfield  {journal} {\bibinfo  {journal} {arXiv preprint arXiv:2308.05793}\ } (\bibinfo {year} {2023})}\BibitemShut {NoStop}%
\bibitem [{\citenamefont {Yang}\ \emph {et~al.}(2017)\citenamefont {Yang}, \citenamefont {Xiang},\ and\ \citenamefont {Wu}}]{yang2017majorana}%
  \BibitemOpen
  \bibfield  {author} {\bibinfo {author} {\bibfnamefont {W.}~\bibnamefont {Yang}}, \bibinfo {author} {\bibfnamefont {T.}~\bibnamefont {Xiang}},\ and\ \bibinfo {author} {\bibfnamefont {C.}~\bibnamefont {Wu}},\ }\bibfield  {title} {\bibinfo {title} {Majorana surface modes of nodal topological pairings in spin-$\frac{3}{2}$ semimetals},\ }\href {https://doi.org/10.1103/PhysRevB.96.144514} {\bibfield  {journal} {\bibinfo  {journal} {Phys. Rev. B}\ }\textbf {\bibinfo {volume} {96}},\ \bibinfo {pages} {144514} (\bibinfo {year} {2017})}\BibitemShut {NoStop}%
\bibitem [{\citenamefont {Roy}\ \emph {et~al.}(2019)\citenamefont {Roy}, \citenamefont {Ghorashi}, \citenamefont {Foster},\ and\ \citenamefont {Nevidomskyy}}]{roy2019topo}%
  \BibitemOpen
  \bibfield  {author} {\bibinfo {author} {\bibfnamefont {B.}~\bibnamefont {Roy}}, \bibinfo {author} {\bibfnamefont {S.~A.~A.}\ \bibnamefont {Ghorashi}}, \bibinfo {author} {\bibfnamefont {M.~S.}\ \bibnamefont {Foster}},\ and\ \bibinfo {author} {\bibfnamefont {A.~H.}\ \bibnamefont {Nevidomskyy}},\ }\bibfield  {title} {\bibinfo {title} {Topological superconductivity of spin-$3/2$ carriers in a three-dimensional doped luttinger semimetal},\ }\href {https://doi.org/10.1103/PhysRevB.99.054505} {\bibfield  {journal} {\bibinfo  {journal} {Phys. Rev. B}\ }\textbf {\bibinfo {volume} {99}},\ \bibinfo {pages} {054505} (\bibinfo {year} {2019})}\BibitemShut {NoStop}%
\bibitem [{\citenamefont {Boettcher}\ and\ \citenamefont {Herbut}(2018)}]{boettcher2018SC}%
  \BibitemOpen
  \bibfield  {author} {\bibinfo {author} {\bibfnamefont {I.}~\bibnamefont {Boettcher}}\ and\ \bibinfo {author} {\bibfnamefont {I.~F.}\ \bibnamefont {Herbut}},\ }\bibfield  {title} {\bibinfo {title} {Unconventional superconductivity in luttinger semimetals: Theory of complex tensor order and the emergence of the uniaxial nematic state},\ }\href {https://doi.org/10.1103/PhysRevLett.120.057002} {\bibfield  {journal} {\bibinfo  {journal} {Phys. Rev. Lett.}\ }\textbf {\bibinfo {volume} {120}},\ \bibinfo {pages} {057002} (\bibinfo {year} {2018})}\BibitemShut {NoStop}%
\bibitem [{\citenamefont {Yu}\ and\ \citenamefont {Liu}(2018)}]{yu2018singlet}%
  \BibitemOpen
  \bibfield  {author} {\bibinfo {author} {\bibfnamefont {J.}~\bibnamefont {Yu}}\ and\ \bibinfo {author} {\bibfnamefont {C.-X.}\ \bibnamefont {Liu}},\ }\bibfield  {title} {\bibinfo {title} {Singlet-quintet mixing in spin-orbit coupled superconductors with $j=\frac{3}{2}$ fermions},\ }\href {https://doi.org/10.1103/PhysRevB.98.104514} {\bibfield  {journal} {\bibinfo  {journal} {Phys. Rev. B}\ }\textbf {\bibinfo {volume} {98}},\ \bibinfo {pages} {104514} (\bibinfo {year} {2018})}\BibitemShut {NoStop}%
\bibitem [{\citenamefont {Venderbos}\ \emph {et~al.}(2018)\citenamefont {Venderbos}, \citenamefont {Savary}, \citenamefont {Ruhman}, \citenamefont {Lee},\ and\ \citenamefont {Fu}}]{venderbos2018pairing}%
  \BibitemOpen
  \bibfield  {author} {\bibinfo {author} {\bibfnamefont {J.~W.~F.}\ \bibnamefont {Venderbos}}, \bibinfo {author} {\bibfnamefont {L.}~\bibnamefont {Savary}}, \bibinfo {author} {\bibfnamefont {J.}~\bibnamefont {Ruhman}}, \bibinfo {author} {\bibfnamefont {P.~A.}\ \bibnamefont {Lee}},\ and\ \bibinfo {author} {\bibfnamefont {L.}~\bibnamefont {Fu}},\ }\bibfield  {title} {\bibinfo {title} {Pairing states of spin-$\frac{3}{2}$ fermions: Symmetry-enforced topological gap functions},\ }\href {https://doi.org/10.1103/PhysRevX.8.011029} {\bibfield  {journal} {\bibinfo  {journal} {Phys. Rev. X}\ }\textbf {\bibinfo {volume} {8}},\ \bibinfo {pages} {011029} (\bibinfo {year} {2018})}\BibitemShut {NoStop}%
\bibitem [{\citenamefont {Zhu}\ \emph {et~al.}(2023)\citenamefont {Zhu}, \citenamefont {Alexandradinata},\ and\ \citenamefont {Hughes}}]{Zhu2023spin}%
  \BibitemOpen
  \bibfield  {author} {\bibinfo {author} {\bibfnamefont {P.}~\bibnamefont {Zhu}}, \bibinfo {author} {\bibfnamefont {A.}~\bibnamefont {Alexandradinata}},\ and\ \bibinfo {author} {\bibfnamefont {T.~L.}\ \bibnamefont {Hughes}},\ }\bibfield  {title} {\bibinfo {title} {$\mathbb{Z}_{2}$ spin hopf insulator: Helical hinge states and returning thouless pump},\ }\href {https://doi.org/10.1103/PhysRevB.107.115159} {\bibfield  {journal} {\bibinfo  {journal} {Phys. Rev. B}\ }\textbf {\bibinfo {volume} {107}},\ \bibinfo {pages} {115159} (\bibinfo {year} {2023})}\BibitemShut {NoStop}%
\bibitem [{\citenamefont {Piotrzkowski}\ \emph {et~al.}(1965)\citenamefont {Piotrzkowski}, \citenamefont {Porowski}, \citenamefont {Dziuba}, \citenamefont {Ginter}, \citenamefont {Giriat},\ and\ \citenamefont {Sosnowski}}]{piotrzkowski1965band}%
  \BibitemOpen
  \bibfield  {author} {\bibinfo {author} {\bibfnamefont {R.}~\bibnamefont {Piotrzkowski}}, \bibinfo {author} {\bibfnamefont {S.}~\bibnamefont {Porowski}}, \bibinfo {author} {\bibfnamefont {Z.}~\bibnamefont {Dziuba}}, \bibinfo {author} {\bibfnamefont {J.}~\bibnamefont {Ginter}}, \bibinfo {author} {\bibfnamefont {W.}~\bibnamefont {Giriat}},\ and\ \bibinfo {author} {\bibfnamefont {L.}~\bibnamefont {Sosnowski}},\ }\bibfield  {title} {\bibinfo {title} {Band structure of hgte},\ }\href {https://doi.org/10.1002/pssb.19650080333} {\bibfield  {journal} {\bibinfo  {journal} {physica status solidi (b)}\ }\textbf {\bibinfo {volume} {8}},\ \bibinfo {pages} {K135} (\bibinfo {year} {1965})}\BibitemShut {NoStop}%
\bibitem [{\citenamefont {Kondo}\ \emph {et~al.}(2015)\citenamefont {Kondo}, \citenamefont {Nakayama}, \citenamefont {Chen}, \citenamefont {Ishikawa}, \citenamefont {Moon}, \citenamefont {Yamamoto}, \citenamefont {Ota}, \citenamefont {Malaeb}, \citenamefont {Kanai}, \citenamefont {Nakashima} \emph {et~al.}}]{kondo2015quadratic}%
  \BibitemOpen
  \bibfield  {author} {\bibinfo {author} {\bibfnamefont {T.}~\bibnamefont {Kondo}}, \bibinfo {author} {\bibfnamefont {M.}~\bibnamefont {Nakayama}}, \bibinfo {author} {\bibfnamefont {R.}~\bibnamefont {Chen}}, \bibinfo {author} {\bibfnamefont {J.}~\bibnamefont {Ishikawa}}, \bibinfo {author} {\bibfnamefont {E.-G.}\ \bibnamefont {Moon}}, \bibinfo {author} {\bibfnamefont {T.}~\bibnamefont {Yamamoto}}, \bibinfo {author} {\bibfnamefont {Y.}~\bibnamefont {Ota}}, \bibinfo {author} {\bibfnamefont {W.}~\bibnamefont {Malaeb}}, \bibinfo {author} {\bibfnamefont {H.}~\bibnamefont {Kanai}}, \bibinfo {author} {\bibfnamefont {Y.}~\bibnamefont {Nakashima}}, \emph {et~al.},\ }\bibfield  {title} {\bibinfo {title} {Quadratic fermi node in a 3d strongly correlated semimetal},\ }\href {https://doi.org/10.1038/ncomms10042} {\bibfield  {journal} {\bibinfo  {journal} {Nature communications}\ }\textbf {\bibinfo {volume} {6}},\ \bibinfo {pages} {10042} (\bibinfo {year} {2015})}\BibitemShut {NoStop}%
\end{thebibliography}%


\begin{thebibliography}{5}%
\makeatletter
\providecommand \@ifxundefined [1]{%
 \@ifx{#1\undefined}
}%
\providecommand \@ifnum [1]{%
 \ifnum #1\expandafter \@firstoftwo
 \else \expandafter \@secondoftwo
 \fi
}%
\providecommand \@ifx [1]{%
 \ifx #1\expandafter \@firstoftwo
 \else \expandafter \@secondoftwo
 \fi
}%
\providecommand \natexlab [1]{#1}%
\providecommand \enquote  [1]{``#1''}%
\providecommand \bibnamefont  [1]{#1}%
\providecommand \bibfnamefont [1]{#1}%
\providecommand \citenamefont [1]{#1}%
\providecommand \href@noop [0]{\@secondoftwo}%
\providecommand \href [0]{\begingroup \@sanitize@url \@href}%
\providecommand \@href[1]{\@@startlink{#1}\@@href}%
\providecommand \@@href[1]{\endgroup#1\@@endlink}%
\providecommand \@sanitize@url [0]{\catcode `\\12\catcode `\$12\catcode `\&12\catcode `\#12\catcode `\^12\catcode `\_12\catcode `\%12\relax}%
\providecommand \@@startlink[1]{}%
\providecommand \@@endlink[0]{}%
\providecommand \url  [0]{\begingroup\@sanitize@url \@url }%
\providecommand \@url [1]{\endgroup\@href {#1}{\urlprefix }}%
\providecommand \urlprefix  [0]{URL }%
\providecommand \Eprint [0]{\href }%
\providecommand \doibase [0]{https://doi.org/}%
\providecommand \selectlanguage [0]{\@gobble}%
\providecommand \bibinfo  [0]{\@secondoftwo}%
\providecommand \bibfield  [0]{\@secondoftwo}%
\providecommand \translation [1]{[#1]}%
\providecommand \BibitemOpen [0]{}%
\providecommand \bibitemStop [0]{}%
\providecommand \bibitemNoStop [0]{.\EOS\space}%
\providecommand \EOS [0]{\spacefactor3000\relax}%
\providecommand \BibitemShut  [1]{\csname bibitem#1\endcsname}%
\let\auto@bib@innerbib\@empty
\bibitem [{\citenamefont {Sun}\ \emph {et~al.}(2018)\citenamefont {Sun}, \citenamefont {Zhang},\ and\ \citenamefont {Bzdu\ifmmode~\check{s}\else \v{s}\fi{}ek}}]{Sun2018conversion}%
  \BibitemOpen
  \bibfield  {author} {\bibinfo {author} {\bibfnamefont {X.-Q.}\ \bibnamefont {Sun}}, \bibinfo {author} {\bibfnamefont {S.-C.}\ \bibnamefont {Zhang}},\ and\ \bibinfo {author} {\bibfnamefont {T.~c.~v.}\ \bibnamefont {Bzdu\ifmmode~\check{s}\else \v{s}\fi{}ek}},\ }\bibfield  {title} {\bibinfo {title} {Conversion rules for weyl points and nodal lines in topological media},\ }\href {https://doi.org/10.1103/PhysRevLett.121.106402} {\bibfield  {journal} {\bibinfo  {journal} {Phys. Rev. Lett.}\ }\textbf {\bibinfo {volume} {121}},\ \bibinfo {pages} {106402} (\bibinfo {year} {2018})}\BibitemShut {NoStop}%
\bibitem [{\citenamefont {Nelson}\ \emph {et~al.}(2022)\citenamefont {Nelson}, \citenamefont {Neupert}, \citenamefont {Alexandradinata},\ and\ \citenamefont {Bzdu\ifmmode~\check{s}\else \v{s}\fi{}ek}}]{Nelson2022delicate}%
  \BibitemOpen
  \bibfield  {author} {\bibinfo {author} {\bibfnamefont {A.}~\bibnamefont {Nelson}}, \bibinfo {author} {\bibfnamefont {T.}~\bibnamefont {Neupert}}, \bibinfo {author} {\bibfnamefont {A.}~\bibnamefont {Alexandradinata}},\ and\ \bibinfo {author} {\bibfnamefont {T.~c.~v.}\ \bibnamefont {Bzdu\ifmmode~\check{s}\else \v{s}\fi{}ek}},\ }\bibfield  {title} {\bibinfo {title} {Delicate topology protected by rotation symmetry: Crystalline hopf insulators and beyond},\ }\href {https://doi.org/10.1103/PhysRevB.106.075124} {\bibfield  {journal} {\bibinfo  {journal} {Phys. Rev. B}\ }\textbf {\bibinfo {volume} {106}},\ \bibinfo {pages} {075124} (\bibinfo {year} {2022})}\BibitemShut {NoStop}%
\bibitem [{\citenamefont {Brydon}\ \emph {et~al.}(2016)\citenamefont {Brydon}, \citenamefont {Wang}, \citenamefont {Weinert},\ and\ \citenamefont {Agterberg}}]{pairing2016brydon}%
  \BibitemOpen
  \bibfield  {author} {\bibinfo {author} {\bibfnamefont {P.~M.~R.}\ \bibnamefont {Brydon}}, \bibinfo {author} {\bibfnamefont {L.}~\bibnamefont {Wang}}, \bibinfo {author} {\bibfnamefont {M.}~\bibnamefont {Weinert}},\ and\ \bibinfo {author} {\bibfnamefont {D.~F.}\ \bibnamefont {Agterberg}},\ }\bibfield  {title} {\bibinfo {title} {Pairing of $j=3/2$ fermions in half-heusler superconductors},\ }\href {https://doi.org/10.1103/PhysRevLett.116.177001} {\bibfield  {journal} {\bibinfo  {journal} {Phys. Rev. Lett.}\ }\textbf {\bibinfo {volume} {116}},\ \bibinfo {pages} {177001} (\bibinfo {year} {2016})}\BibitemShut {NoStop}%
\bibitem [{\citenamefont {Kim}\ \emph {et~al.}(2018)\citenamefont {Kim}, \citenamefont {Wang}, \citenamefont {Nakajima}, \citenamefont {Hu}, \citenamefont {Ziemak}, \citenamefont {Syers}, \citenamefont {Wang}, \citenamefont {Hodovanets}, \citenamefont {Denlinger}, \citenamefont {Brydon} \emph {et~al.}}]{kim2018beyond}%
  \BibitemOpen
  \bibfield  {author} {\bibinfo {author} {\bibfnamefont {H.}~\bibnamefont {Kim}}, \bibinfo {author} {\bibfnamefont {K.}~\bibnamefont {Wang}}, \bibinfo {author} {\bibfnamefont {Y.}~\bibnamefont {Nakajima}}, \bibinfo {author} {\bibfnamefont {R.}~\bibnamefont {Hu}}, \bibinfo {author} {\bibfnamefont {S.}~\bibnamefont {Ziemak}}, \bibinfo {author} {\bibfnamefont {P.}~\bibnamefont {Syers}}, \bibinfo {author} {\bibfnamefont {L.}~\bibnamefont {Wang}}, \bibinfo {author} {\bibfnamefont {H.}~\bibnamefont {Hodovanets}}, \bibinfo {author} {\bibfnamefont {J.~D.}\ \bibnamefont {Denlinger}}, \bibinfo {author} {\bibfnamefont {P.~M.}\ \bibnamefont {Brydon}}, \emph {et~al.},\ }\bibfield  {title} {\bibinfo {title} {Beyond triplet: Unconventional superconductivity in a spin-3/2 topological semimetal},\ }\href {https://www.science.org/doi/10.1126/sciadv.aao4513} {\bibfield  {journal} {\bibinfo  {journal} {Science advances}\ }\textbf {\bibinfo {volume} {4}},\ \bibinfo {pages} {eaao4513} (\bibinfo {year} {2018})}\BibitemShut
  {NoStop}%
\bibitem [{\citenamefont {Ishihara}\ \emph {et~al.}(2021)\citenamefont {Ishihara}, \citenamefont {Takenaka}, \citenamefont {Miao}, \citenamefont {Mizukami}, \citenamefont {Hashimoto}, \citenamefont {Yamashita}, \citenamefont {Konczykowski}, \citenamefont {Masuki}, \citenamefont {Hirayama}, \citenamefont {Nomoto}, \citenamefont {Arita}, \citenamefont {Pavlosiuk}, \citenamefont {Wi\ifmmode~\acute{s}\else \'{s}\fi{}niewski}, \citenamefont {Kaczorowski},\ and\ \citenamefont {Shibauchi}}]{tuning2021ishihara}%
  \BibitemOpen
  \bibfield  {author} {\bibinfo {author} {\bibfnamefont {K.}~\bibnamefont {Ishihara}}, \bibinfo {author} {\bibfnamefont {T.}~\bibnamefont {Takenaka}}, \bibinfo {author} {\bibfnamefont {Y.}~\bibnamefont {Miao}}, \bibinfo {author} {\bibfnamefont {Y.}~\bibnamefont {Mizukami}}, \bibinfo {author} {\bibfnamefont {K.}~\bibnamefont {Hashimoto}}, \bibinfo {author} {\bibfnamefont {M.}~\bibnamefont {Yamashita}}, \bibinfo {author} {\bibfnamefont {M.}~\bibnamefont {Konczykowski}}, \bibinfo {author} {\bibfnamefont {R.}~\bibnamefont {Masuki}}, \bibinfo {author} {\bibfnamefont {M.}~\bibnamefont {Hirayama}}, \bibinfo {author} {\bibfnamefont {T.}~\bibnamefont {Nomoto}}, \bibinfo {author} {\bibfnamefont {R.}~\bibnamefont {Arita}}, \bibinfo {author} {\bibfnamefont {O.}~\bibnamefont {Pavlosiuk}}, \bibinfo {author} {\bibfnamefont {P.}~\bibnamefont {Wi\ifmmode~\acute{s}\else \'{s}\fi{}niewski}}, \bibinfo {author} {\bibfnamefont {D.}~\bibnamefont {Kaczorowski}},\ and\ \bibinfo {author} {\bibfnamefont {T.}~\bibnamefont {Shibauchi}},\
  }\bibfield  {title} {\bibinfo {title} {Tuning the parity mixing of singlet-septet pairing in a half-heusler superconductor},\ }\href {https://doi.org/10.1103/PhysRevX.11.041048} {\bibfield  {journal} {\bibinfo  {journal} {Phys. Rev. X}\ }\textbf {\bibinfo {volume} {11}},\ \bibinfo {pages} {041048} (\bibinfo {year} {2021})}\BibitemShut {NoStop}%
\end{thebibliography}%




\end{document}


\title{Supplemental Material for ``Dipole-Obstructed Cooper Pairing: Theory and Application to $j=3/2$ Superconductors"}

\author{Penghao Zhu} 
\email{zhu.3711@osu.edu}
\affiliation{Department of Physics, The Ohio State University, Columbus, OH 43210, USA}
\author{Rui-Xing Zhang}
\email{ruixing@utk.edu}
\affiliation{Department of Physics and Astronomy, The University of Tennessee, Knoxville, TN 37996, USA}
\affiliation{Department of Materials Science and Engineering, The University of Tennessee, Knoxville, TN 37996, USA}


\maketitle
\tableofcontents

\setcounter{section}{0}
\setcounter{figure}{0}
\setcounter{equation}{0}
\renewcommand{\thefigure}{S\arabic{figure}}
\renewcommand{\theequation}{S\arabic{equation}}
\renewcommand{\thesection}{S\arabic{section}}
\onecolumngrid



\section{Proof of Chern-vorticity theorem}

\begin{figure}[h]
    \centering
    \includegraphics[width=0.6\columnwidth]{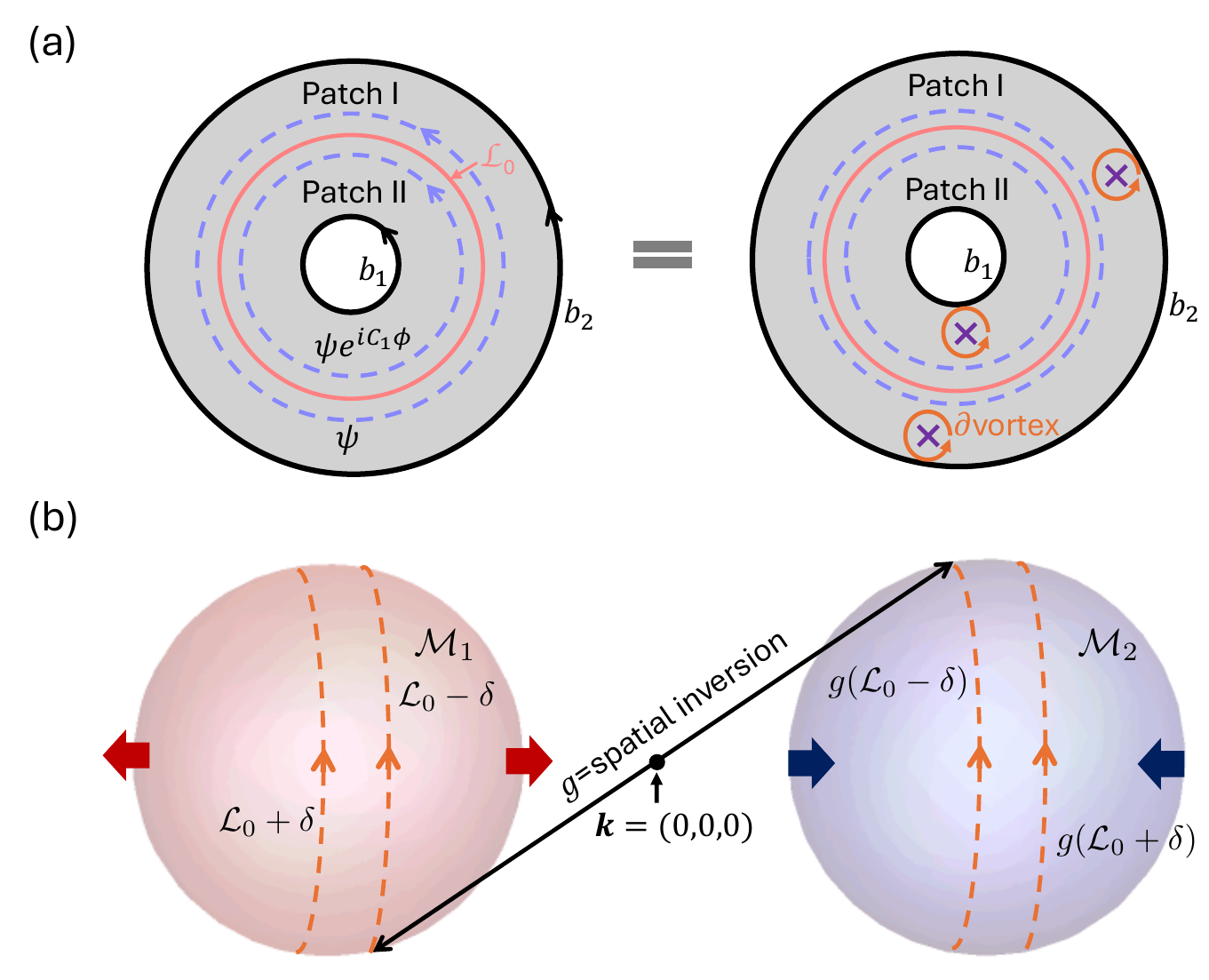}
    \caption{(a) Illustration of the two gauge patches over an annulus geometry, and the paths for the line integrals. The purple crosses represent vortex of $\mathcal{O}(\mathbf{k}_{1}
    )$. The inner (outer) blue dashed circle represents $\mathcal{L}_{0}+\delta$ ($\mathcal{L}_{0}-\delta$). (b) Illustration of the orientation of the Berry flux over $\mathcal{M}_2$ associated with the path integral in Eq.~\eqref{eq:integral} when $g$ is an spatial inversion symmetry. The red and blue arrows denote the positive direction of the Berry flux.}
    \label{fig:chern-vorticity}
\end{figure}

We prove the Chern-vorticity theorem for 
\begin{equation}
    {\cal O}({\bf k}_1) = \bra{\psi_{1}(\mathbf{k}_1)}\hat{O}\ket{\psi_{2}(\mathbf{k}_{2})} = |{\cal O}({\bf k}_1)|e^{i \varphi({\bf k}_1)}.
\end{equation}
by constructing a vector:
\begin{equation}
\label{eq:vector}
\mathbf{S}=\partial_{\mathbf{k}_{1}}\varphi(\mathbf{k}_{1})- \mathbf{A}_{1}(\mathbf{k}_{1})+g^{T}\mathbf{A}_{2}(\mathbf{k}_{2}),
\end{equation}
where
\begin{equation}
\label{eq:BC}
\mathbf{A}_{a}(\mathbf{k}_{a})=i\langle \psi_{a}(\mathbf{k}_{a})|\partial_{\mathbf{k}_{a}}\psi_{a}(\mathbf{k}_{a})\rangle, \ a=1,2,
\end{equation}
is the Berry connection. Importantly, vector $\mathbf{S}$ is invariant under the $U(1)$ gauge transformation $|\psi_{a}(\mathbf{k}_{a})\rangle\rightarrow e^{i\alpha_{a}(\mathbf{k}_{a})}|\psi_{a}(\mathbf{k}_{a})\rangle$ with $a=1,2$.
If there is a nonzero Chern-number of $\ket{\psi_{1}(\bf k_{1})}$ ($\ket{\psi_{2}(\bf k_{2})}$) on effectively closed $\mathcal{M}_{1}$ ($\mathcal{M}_{2}$), then $\mathbf{A}_{1}(\mathbf{k}_1)$ ($\mathbf{A}_{2}(\mathbf{k}_2)$) cannot be globally smooth over the entire manifold. Without loss of generality, one can have two locally smooth gauge patches over the manifold as illustrated for an annulus geometry in Fig.~\ref{fig:chern-vorticity}(a). Then, across the boundary (i.e., $\mathcal{L}_{0}$ in Fig.~\ref{fig:chern-vorticity}(a)) of the two patches, there are singularities of $\mathbf{A}_{1}(\mathbf{k}_1)$ and $\mathbf{A}_{2}(\mathbf{k}_2)$ such that
\begin{equation}
\label{eq:integral}
\begin{aligned}
&\lim_{\delta\rightarrow 0}\left(\int_{\mathcal{L}_{0}+\delta}-\int_{\mathcal{L}_{0}-\delta}\right)\mathbf{A}_{1}(\mathbf{k}_{1})\cdot d\mathbf{k}_{1}=2\pi \mathcal{C}_{1},
\\
& \lim_{\delta\rightarrow 0}\left(\int_{g(\mathcal{L}_{0}+\delta)+\mathbf{t}}-\int_{g(\mathcal{L}_{0}-\delta)+\mathbf{t}}\right)\mathbf{A}_{2}(\mathbf{k}_{2})\cdot d\mathbf{k}_{2}=2\pi (\det g) \mathcal{C}_{2}.
\end{aligned}
\end{equation}
Here, $\det g $ shows up in the second equation above because we have defined $\mathcal{C}_{1}$ and $\mathcal{C}_{2}$ over $\mathcal{M}_{1}$ and $\mathcal{M}_{2}$ under the same orientation-choice, e.g., for a 2D closed surface in 3D, we choose the normal vectors of $\mathcal{M}_{1}$ and $\mathcal{M}_{2}$ both to point outward from their enclosed region. If $g$ is an improper transformation such as inversion, $\det g=-1$ and the path integral $\lim_{\delta\rightarrow 0}\left(\int_{g(\mathcal{L}_{0}+\delta)+\mathbf{t}}-\int_{g(\mathcal{L}_{0}-\delta)+\mathbf{t}}\right)\mathbf{A}_{2}(\mathbf{k}_{2})\cdot d\mathbf{k}_{2}$ will correspond to the Chern number defined with respect to normal vectors pointing inwards the enclosed region as illustrated in Fig.~\ref{fig:chern-vorticity}(b), which is opposite to $\mathcal{C}_{2}$.

In contrast to the Berry connection, $\mathbf{S}$ is gauge invariant, and thus is continuous across the gauge patch boundary, i.e., $\lim_{\delta\rightarrow 0}\left(\int_{\mathcal{L}_{0}+\delta}-\int_{\mathcal{L}_{0}-\delta}\right)\mathbf{S}\cdot d\mathbf{k}_{1}=0$. Consequently, $\partial_{\mathbf{k}_{1}}\varphi(\mathbf{k}_{1})$ must have its own singularities (i.e., vortices) to cancel or balance those from the Berry connection, which leads to
\begin{equation}
\label{eq:integral1}
\begin{aligned}
&\lim_{\delta\rightarrow 0}\left(\int_{\mathcal{L}_{0}+\delta}-\int_{\mathcal{L}_{0}-\delta}\right)\partial_{\mathbf{k}_{1}}\varphi\cdot d\mathbf{k}_{1}=\lim_{\delta\rightarrow 0}\left(\int_{\mathcal{L}_{0}+\delta}-\int_{\mathcal{L}_{0}-\delta}\right)[\mathbf{A}_{1}(\mathbf{k}_{1})-g^{T}\mathbf{A}_{2}(\mathbf{k}_{2})]\cdot d\mathbf{k}_{1}
\\
&=\lim_{\delta\rightarrow 0}\left(\int_{\mathcal{L}_{0}+\delta}-\int_{\mathcal{L}_{0}-\delta}\right)\mathbf{A}_{1}(\mathbf{k}_{1})\cdot d\mathbf{k}_{1}-\lim_{\delta\rightarrow 0}\left(\int_{g(\mathcal{L}_{0}+\delta)+\mathbf{t}}-\int_{g(\mathcal{L}_{0}-\delta)+\mathbf{t}}\right)\mathbf{A}_{2}(\mathbf{k}_{2})\cdot d\mathbf{k}_{2}
\\
&=2\pi(\mathcal{C}_{1}-(\det{g})\mathcal{C}_{2}).
\end{aligned}
\end{equation}

As illustrated in Fig.~\ref{fig:chern-vorticity}(a), the net vorticity of $\mathcal{O}(\mathbf{k}_{1})$, defined as $\nu=\sum_{\text{vortex}}\oint_{\partial\text{vortex}}\partial_{\mathbf{k}_{1}}\varphi\cdot d\mathbf{k}_{1}$, can be calculated through line integrals over $\mathcal{L}_{0}\pm\delta$ and $b_{1,2}$:
\begin{equation}
\label{eq:netv}
\begin{aligned}
\nu&=\frac{1}{2\pi}\left(\oint_{b_2}-\oint_{b_1}\right)\partial_{\mathbf{k}_{1}}\varphi\cdot d\mathbf{k}_{1}+\frac{1}{2\pi}\lim_{\delta\rightarrow 0}\left(\int_{\mathcal{L}_{0}+\delta}-\int_{\mathcal{L}_{0}-\delta}\right)\partial_{\mathbf{k}_{1}}\varphi\cdot d\mathbf{k}_{1}
\\
&=\mathcal{C}_{1}-(\det g)\mathcal{C}_{2}+\frac{1}{2\pi}\left(\oint_{b_2}-\oint_{b_1}\right)\partial_{\mathbf{k}_{1}}\varphi\cdot d\mathbf{k}_{1}
\end{aligned}
\end{equation}
Rewriting $1/(2\pi)\left(\oint_{b_2}+\oint_{b_1}\right)\partial_{\mathbf{k}_{1}}\varphi\cdot d\mathbf{k}_{1}$ formally as $1/(2\pi)\sum_{b}\oint_{b}\partial_{\mathbf{k}_{1}}\varphi\cdot d\mathbf{k}_{1}$ results in the Chern-vorticity theorem depicted in Eq.~(2) in the main text.
 
For SC systems where $\ket{\psi_{2}(\mathbf{k}_2)}=\ket{\xi_{2}^{(e)\star}(\mathbf{k}_2)}$, 
$$\mathbf{A}_{2}(\mathbf{k}_{2})=i\bra{\xi^{(e)\star}_{2}(\mathbf{k}_{2})}\partial_{\mathbf{k}_{2}}\xi^{(e)\star}_{2}(\mathbf{k}_{2})\rangle=-i\bra{\xi^{(e)}_{2}(\mathbf{k}_{2})}\partial_{\mathbf{k}_{2}}\xi^{(e)}_{2}(\mathbf{k}_{2})\rangle,$$ 
and thus 
$$\mathcal{C}_{2}\equiv \lim_{\delta\rightarrow 0}\left(\int_{-\mathcal{L}_{0}+\delta}-\int_{-\mathcal{L}_{0}-\delta}\right) \left[-i\bra{\xi^{(e)}_{2}(\mathbf{k}_{2})}\partial_{\mathbf{k}_{2}}\xi^{(e)}_{2}(\mathbf{k}_{2})\rangle\right]\cdot d\mathbf{k}_{2},$$ which is opposite to the Chern number of electronic Bloch states $\ket{\xi^{(e)}_{2}(\mathbf{k}_{2})}$.

\section{Mirror symmetry and quantized Berry-dipole flux in two-band tight-binding models}

In this part, we prove that the mirror symmetry can quantize Berry-dipole flux over the FS~\cite{Sun2018conversion,Nelson2022delicate} for general two-band models. Let us consider a general two-band Hamiltonian with $H({\bf k})=\mathbf{h}(\mathbf{k})\cdot\boldsymbol{\sigma}$ that respects a $\hat{z}$-directional mirror symmetry $M_{z}=i\sigma_{z}$. Here $\boldsymbol{\sigma}=(\sigma_{x},\sigma_{y},\sigma_{z})$ are Pauli matrices. Crucially, $M_z$ requires $h_{x,y}(-k_z)=-h_{x,y}(k_z)$ and $h_{z}(-k_z)=h_{z}(k_z)$. For a spherical-like FS ${\cal M}$ enclosing $\Gamma=(0,0,0)$, we denote ${\cal M}_0$ as the 1D ``equator" at $k_z=0$ [i.e., the black loop in Fig.~(1))(b) in the main text] and ${\cal M}_\pm$ as the FS patches with $k_z>0$ and $k_z<0$, respectively. Since $M_z$ requires $H(k_z=0)$ to be diagonal, either valence or conduction states along ${\cal M}_0$ shall be identical, up to an unimportant $U(1)$ phase. In other words, by choosing a proper gauge to eliminate this $U(1)$ phase, the 1D equator ${\cal M}_0$ is topologically equivalent to a 0D point, and both ${\cal M}_{\pm}$ are then effectively closed 2D manifolds, over which the Berry flux must be quantized. Please note that this quantization condition does not generally apply to $N$-band models with $N>2$, which may require additional symmetry protection.

\section{Cooper pairings of Berry-dipole Fermi surface}

The general pairing matrix for a Berry-dipole system is given by
\begin{equation}
\label{eq:bdpairing}
\Delta(\mathbf{k})=d_{x}(\mathbf{k})\sigma_{x}+d_{y}(\mathbf{k})\sigma_{y}+d_{+}\sigma_{+}+d_{-}\sigma_{-},
\end{equation} 
where $\sigma_{\pm}=(\sigma_{z}\pm \sigma_{0})/2$.
The Fermi statistics requires $\Delta(\mathbf{k})=-\Delta^{T}(-\mathbf{k})$, and thus $d_{x,\pm}(\mathbf{k})$ ($d_{y}(\mathbf{k})$) are (is) odd (even) in $\mathbf{k}$. 

Note that when $\Sigma=0$ in Eq.~(3) of the main text, $\mathcal{M}_{\pm}$ are hemispheres and the only boundary for both of them is the equator. This corresponds to the case where $b_{1}$ in Fig.~\ref{fig:chern-vorticity}(a) shrinks into a point.

\subsection{Projection onto FS}
Utilizing the gauge patterns and the wavefunction specified in the main text, we can derive a simple formula for $\Delta_{\text{eff}}=\bra{\xi^{(e)\star}_{-}(-\mathbf{k})}\Delta(\mathbf{k})\ket{\xi^{(e)}_{+}(\mathbf{k})}$ in the two gauge patches:
\begin{equation}
\label{eq:deltaeff}
\begin{aligned}
    &\Delta_{\text{eff},I}= d_x e^{i\phi}\sin 2\theta-(d_{+}e^{2i\phi}\cos^2\theta+d_{-}\sin^2\theta),
    \\
    &\Delta_{\text{eff},II}= d_x e^{-i\phi}\sin 2\theta-(d_{+}\cos^2\theta+d_{-}e^{-2i\phi}\sin^2\theta).
\end{aligned}
\end{equation}
The consequence of each pairing channel is summarized as follows:
\begin{itemize}
    \item $d_{x}$ channel has zeros at $\theta=0,\pi/2,\pi$. If we count the vorticity on $\mathcal{M}_{+}$ following Eq.~\eqref{eq:netv}, we will find a net vorticity $-2$: Phase winding $2\pi$ of $\Delta_{\text{eff},I}$ indicates a vorticity $-1$ in patch I, and the $-2\pi$ of $\Delta_{\text{eff},II}$ indicates a vorticity $-1$ in patch II. This is because there is a sign difference between the integrals over $\mathcal{L}_{0}+\delta$ and $\mathcal{L}_{0}-\delta$ in the definition of the vorticity [c.f. Eq.~\eqref{eq:netv}]. Please note that the above counting of $\nu=2$ has NOT taken into account the explicit expression of $d_{x}({\bf k})$, which can update $\nu$ via the boundary integral ${\cal I}_\varphi$. 
    \item $d_{y}$ channel vanishes uniformly when projecting onto the Fermi surface.
    \item $d_{+}$ and $d_{-}$ channels both manifest zeros at $\theta=\pi/2$ (equator) and $\theta=0,\pi$ (poles), of which the vorticity can be counted in the same way as that of $d_x$ pairing.
\end{itemize}

For a general $\Delta({\bf k})$ that consists of non-zero $d_{x}$ and $d_{\pm}$ components, the pairing zeros can appear at arbitrary positions. As an example, we study $d_{x}=k_{z}$ and $d_{\pm}=k_{\mp}$, where $k_{\pm}=k_{x}\pm i k_{y}$. Substituting these into Eq.~\eqref{eq:deltaeff}, we solve
\begin{equation}
\Delta_{\text{eff},I}=0=e^{i\phi}\cos\theta\sin2\theta-e^{i\phi}\sin\theta\cos^2\theta-e^{i\phi}\sin^3\theta,
\end{equation}
which gives us zeros at $\theta=0,\pi/4,3\pi/4,\pi$. This can be understood in the following way: (1) When $d_{x}$ is the only nonzero term, there are zeros at the poles and the equator; (2) When $d_{\pm}$ are turned on, the zeros at the equator are separated into two loops at general latitude. In this case, the boundary term in Eq.(2) is determined by the phase winding of $d_{-}$ over the equator, which contributes $+1$ to $\nu$ according to Eq.~\eqref{eq:netv}. Since the $\mathcal{C}_{1}-(\det g)\mathcal{C}_{2}=-2$ for $\mu<0$, the net vorticity in this case is $-1$. Importantly, the nodal points observed at poles in this example are the origin of those observed in LSM, as discussed in the main text.

\subsection{Symmetry analysis and robust nodal points under certain channels}

Here, we first examine the $C_{4h}$ point symmetry group as specified in the main text and discuss the constraints imposed on pairings (i.e. $d_{x,y,\pm}$ in Eq.~\eqref{eq:bdpairing}). The character table for $C_{4h}$ is shown in Table.~\ref{tab:C_4h}. In SC systems, we call pairing $\Delta(\mathbf{k})$ is an irreducible representation (irrep)/symmetric channel of $C_{4h}$ group if 
\begin{equation}
\label{eq:rep}
\mathring{\mathcal{G}}\Delta(\mathbf{k})\mathring{\mathcal{G}}^{T}=\alpha_{\mathcal{G}}\Delta(\check{\mathcal{G}
}\mathbf{k}), \ \forall \mathcal{G}\in C_{4h},
\end{equation}
where $\mathring{\mathcal{G}}$ ($\check{\mathcal{G}
}$) is the representation of $\mathcal{G}$ in the Hilbert space of electrons (momentum space), and  $\alpha_{\mathcal{G}}$ is the character of $\mathcal{G}$ given by the body of Table.~\ref{tab:C_4h}. Different irreps/symmetric channels correspond to different representations of symmetry operators (denoted as $\acute{\mathcal{G}}$) for the BdG Hamiltonian, i.e.,
\begin{equation}
\acute{\mathcal{G}}=\begin{pmatrix}
\mathring{\mathcal{G}} & 
\\
  & \alpha_{\mathcal{G}}\mathring{\mathcal{G}}^{\star}
\end{pmatrix}
\end{equation}
First, for irreps/symmetric channels $A_{g},B_{g},^{1}E_{g},^{2}E_{g}$ ($A_{u},B_{u},^{1}E_{u},^{2}E_{u}$), $d_{x,y,\pm}$ are enforced to be even (odd) in $\mathbf{k}$ because $P \sigma_{x,y,\pm}P^{T}=\sigma_{x,y,\pm}$. Combining with the requirements of Fermi statistics, we conclude that $d_{x,\pm}$ ($d_{y}$) vanish(es) for irreps/symmetric channels $A_{g},B_{g},^{1}E_{g},^{2}E_{g}$ ($A_{u},B_{u},^{1}E_{u},^{2}E_{u}$). Since the $d_{y}$ term vanishes uniformly when projecting onto the FS, we should next focus on irreps/symmetric channels $A_{u},B_{u},^{1}E_{u},^{2}E_{u}$ and consider the constraints from $C_{4}$. In our case, $C_{4}$ acts on Pauli matrices as 
\begin{equation}
\label{eq:c4pauli}
C_4 \sigma_{\pm} C_4^T = \pm i \sigma_\pm,\ C_4 \sigma_{x,y} C_4^T = - \sigma_{x,y}.
\end{equation}
Consequently, $d_{x/\pm}(\mathbf{k})=\beta_{x/\pm}d_{x/\pm}(C_{4}\mathbf{k})$, here $\beta_{x/\pm}$ is determined by the character of $C_{4}$: $\beta_{x}=-\alpha_{C_{4}},\beta_{\pm}=\mp i\alpha_{C_{4}}$. Importantly, (i)in irreps/symmetric channels where $\beta_{x,\pm}\neq 1$, $d_{x,\pm}$ must vanish at the poles of the FS which are $C_4$ fixed points; and (ii)  in irreps/symmetric channels where $\beta_{x,\pm}=\pm 1$, $d_{x,\pm}$ are even in $(k_{x},k_{y})$ and thus must be odd in $k_{z}$. Consequently, $d_{x,\pm}$ vanish at the equator where $k_{z}$=0.  Following points (i-ii), we can conclude that for  irreps $A_{u}$ and $B_{u}$, $d_{x}$ vanishes when $k_{z}=0$, and $d_{\pm}$ vanish at the poles ($k_{x}=k_{y}=0$) of the FS. In contrast, for irreps $^{1,2}E_{u}$, $d_{x}$ vanishes at poles, and $d_{\pm}$ vanish when $k_{z}=0$ plane.

For $|\mu|>|\Sigma|$ where the FS of $h_{0}$ is always singly connected and sphere-like, the normal states are identically $(1,0)^{T}$ ($(0,1)^T$) for $\mu>0$ ($\mu<0$) over the equator of the FS at $k_{z}=0$ plane. Hence,  $\Delta_{\text{eff}}=d_{+} (d_{-})$ for $\mu>0$ ($\mu<0$). Consequently, the irreps $^{1,2}E_{u}$ with vanishing $d_{\pm}$ corresponds guarantees  the boundary term in Eq.~(2) of the main text to be zero. However, the irrep $A_{u}$ ($B_{u}$) can have nonzero $d_{\pm}$ with $\beta_{\pm}=\mp i (\pm i)$. Considering the constraints imposed by $C_{4}$ rotation symmetry, a smooth pairing factor, $d_{\pm}$, in irrep $A_{u}$ ($B_{u}$) will generally takes the form $d_{\pm}=\sum_{m,n}f_{m,n}k_{+}^{m}k_{-}^{n}$ with $m-n=4p\pm 1$ ($m-n=4p\mp 1$) and $m,n,p$ to be integers. Assuming $k_{x}$ and $k_{y}$ are small, which is automatically valid for our discussion on $\mathbf{k}\cdot\mathbf{p}$ models, the phase winding of $d_{\pm}$ will be determined by the leading term with smallest $m+n$, and generally takes the value $4p \pm 1$ ( $4p \mp 1$ ) times of $2\pi$ where $p$ is an integer. Therefore, \textit{the boundary term in Eq.~(2) of the main text either has no contribution or contributes an odd number to $\nu$}. 

\begin{figure}[t]
    \centering
    \includegraphics[width=0.6\columnwidth]{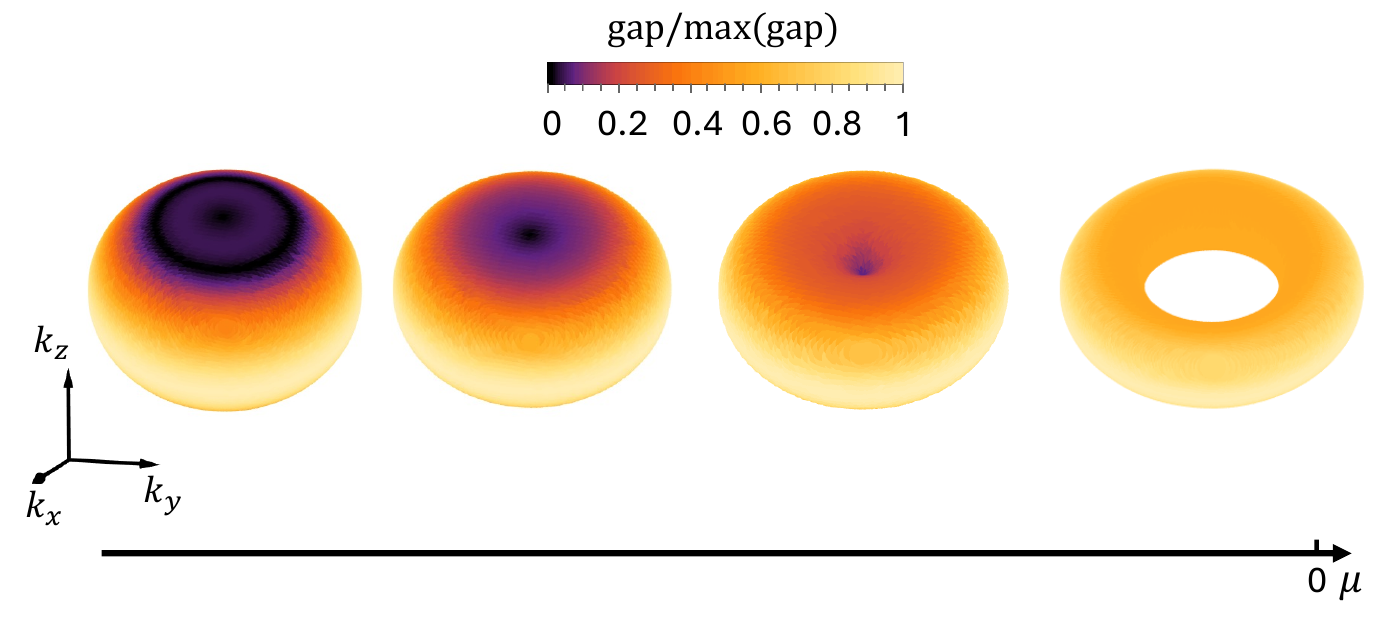}
    \caption{Superconducting gap over the FS for $\Sigma=-0.1$ and pairing $k_{z}\sigma_{x}+k_{-}\sigma_{+}+k_{+}\sigma_{-}$ belongs to the $B_{u}$ irrep, with various chemical potentials $\mu=-0.25,-0.15,-0.1,-0.05$.}
    \label{fig:exception}
\end{figure}

For $|\mu|<|\Sigma|$, the WSM case ($\Sigma>0$) has two spherical FSs with no boundaries. Therefore, the inter-FS pairings are always nodal due to the Chern-vorticity theorem. However,the NLSM case ($\Sigma<0$) has its FS to be a torus, where each of $\mathcal{M}_{+}$ and  $\mathcal{M}_{-}$ has two boundaries. Following the discussion in last paragraph, in $A_{u}$ and $B_{u}$ irreps, each boundary of $\mathcal{M}_{+}$ ($\mathcal{M}_{-}$) can contribute an odd number to the net vorticity $\nu$. Therefore, the net vorticity can be zero and the pairings are not necessarily nodal. Indeed, as shown in Fig.~\ref{fig:exception}, when we have pairing $k_{z}\sigma_{x}+k_{-}\sigma_{+}+k_{+}\sigma_{-}$, the BdG Hamiltonian is totally gapped when $\Sigma<0$ and $|\mu|<|\Sigma|$. In conclusion, the $\Sigma<0$ and $|\mu|<|\Sigma|$ case with pairing channels $A_{u}$ and/or $B_{u}$ irreps of $C_{4h}$ is the only case where the SC pairing over a singly-connected Berry-dipole FS is not topologically obstructed.

\begin{table}
    \centering
    \begin{tabular}{ccccccccc} \hline 
         & $A_g$ & $B_g$ & $^1E_g$  & $^2E_g$ & $A_u$ & $B_u$ & $^1E_u$ & $^2E_u$ \\  \hline 
        $C_4$ &  $1$ & $-1$ & $-i$ & $+i$ & $1$ & $-1$ & $-i$ & $+i$ \\
         $P$ & $1$ & $1$ & $1$ & $1$ & $-1$ & $-1$ & $-1$ & $-1$ \\ \hline
    \end{tabular}
    \caption{Character Table for $C_{4h}$.}
    \label{tab:C_4h}
\end{table}

\section{Annihilation of nodal points upon varying Fermi level in normal Weyl semimetals}

\begin{figure}[h]
    \centering
    \includegraphics[width=0.5\columnwidth]{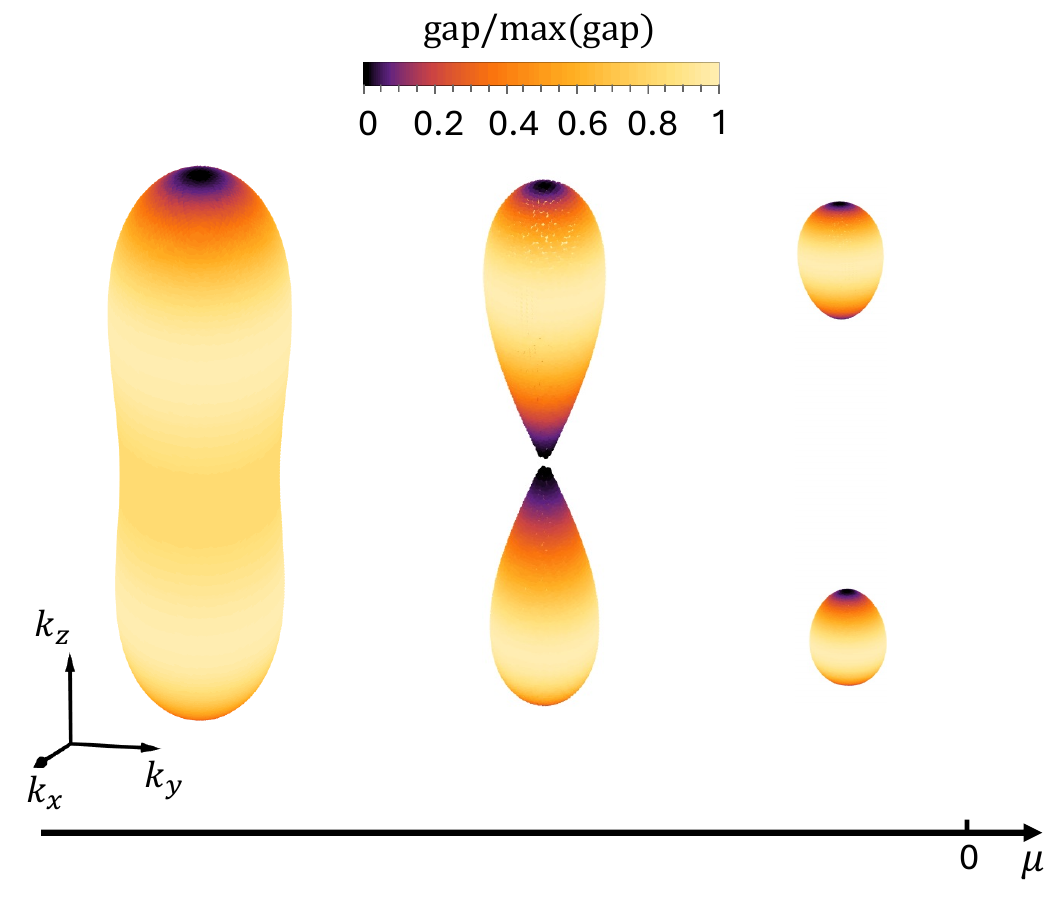}
    \caption{Superconducting gap over the FS for $H_{\text{wsm}}(\mathbf{k})$ with $m=0.1$ and various chemical potentials $\mu=-0.2,-0.1,-0.05$. In this case, the annihilation of BdG Weyl nodes by decreasing $\mu$ leads to a full energy gap.}
    \label{fig:wsm}
\end{figure}

Here, we study a minimal, inversion symmetric model for the Weyl semimetal:
\begin{equation}
\label{eq:hwsm}
h_{\text{wsm}}(\mathbf{k})=k_{x}\sigma_{x}+k_{y}\sigma_{y}+(m-1/2(k_{x}^2+k_{y}^2+k_{z}^2))\sigma_{z},
\end{equation}
and the corresponding BdG Hamiltonian:
\begin{equation}
\label{eq:hwsmbdg}
H_{\text{wsm}}(\mathbf{k})=\begin{pmatrix}
    h_{\text{wsm}}(\mathbf{k}) & \Delta(\mathbf{k})
    \\
    \Delta^{\dag}(\mathbf{k})  &  -h^{T}_{\text{wsm}}(-\mathbf{k})
\end{pmatrix}.
\end{equation}
To make a direct comparison with the Berry-dipole WSM discussed in the main text, we focus on the same pairing $\Delta(\mathbf{k})=(k_{x}+i k_{y})\sigma_{x}$, and calculate the SC gap over the FS for various chemical potentials. As shown in Fig.~\ref{fig:wsm}, the two nodal points merge and annihilate each other upon decreasing $\mu$ ($\mu<0$). This is fundamentally distinct from the node-loop conversion found for the Berry-dipole FS in the main text.

\section{Projection of inter-sector pairing onto FS and symmetry analysis}
\label{sec:inter-sector}
In this section, we analyze the nodal behaviors of inter-sector pairings in Dirac dipole semimetals, consisting of two copies of Berry-dipole semimetals ($h_{0}$ in the main text) in opposite orbital sectors ($l=\pm 1$):
\begin{equation}
\label{eq:dd}
h_{DD}(\mathbf{k})=\begin{pmatrix}
h_{0}(\mathbf{k}) & 0
\\
0 & -h_{0}(\mathbf{k})
\end{pmatrix}.
\end{equation}
$h_{DD}$ has the $C_{4h}\times \mathcal{T}$ symmetry as mentioned in the main text. To simplify the analysis, we focus on $\Sigma=0$ and $\mu<0$ in the following.

Now the pairing matrix is a $4\times 4$ matrix, which can be formally described by
\begin{eqnarray}
    \Delta = \Delta_{\text{intra}} + \Delta_{\text{inter}} = \begin{pmatrix}
        \Delta_{++} & 0 \\
        0 & \Delta_{--} \\
    \end{pmatrix} + \begin{pmatrix}
        0 & \Delta_{+-} \\
        \Delta_{-+} & 0 \\
    \end{pmatrix}.
\end{eqnarray}
The nodal structures of $\Delta_{\text{intra}}$ trivially duplicate those of Berry-dipole semimetals, and here we focus on $\Delta_{\text{inter}}$ that generally takes the form
\begin{equation}
\label{eq:interspinpairing}
\Delta_{\text{inter}}(\mathbf{k})=d_{i,j}(\mathbf{k})\tau_{i}\sigma_{j}, \ i=x,y \ \operatorname{and} \ j=0,x,y,z,
\end{equation}
where $\sigma_{x,y,z}$ ($\tau_{x,y,z}$) represent Pauli matrices for spin (orbital).

\subsection{Projection onto FS}
Under the same gauge used for Berry-dipole semimetals, the eigenstate wavefunctions of $h_{DD}$ in two gauge patches are
\begin{eqnarray}
    && |\xi^{(e)}_{l=+1,I} \rangle =  (-\cos\theta e^{-i\phi}, \sin \theta,0,0)^T,  \ |\xi^{(e)}_{l=-1,I} \rangle = -(0,0,\sin \theta, \cos\theta e^{i\phi})^T,    \nonumber \\
     && |\xi^{(e)}_{l=+1,II} \rangle =  (-\cos\theta, \sin \theta e^{i\phi},0,0)^T,  \ |\xi^{(e)}_{l=-1,II} \rangle = (0,0,\sin \theta e^{-i\phi}, \cos\theta)^T.
    \label{eq:DiracDipole_eigenstate}
\end{eqnarray}
A direct calculation of $\Delta_{\text{eff}}^{\text{inter}}=\bra{\xi^{(e)}_{l=+1}(\mathbf{k})}\Delta_{\text{inter}}\ket{\xi^{(e)\star}_{l=-1}(-\mathbf{k})}$ in the two gauge patches gives
\begin{equation}
\begin{aligned}
 \Delta^{\text{inter}}_{\text{eff},I}= \Delta^{\text{inter}}_{\text{eff},II} &= [-d_{y,y} + \cos 2\theta d_{x,x} + \sin 2\theta (d_{x,z} \cos \phi + d_{y,0} \sin \phi)] \sigma_x \nonumber \\
    & + [d_{x,y} + \cos 2\theta d_{y,x} + \sin 2\theta (d_{y,z}\cos \phi - d_{x,0}\sin \phi)] \sigma_y. 
\end{aligned}
\end{equation}
Let us first consider each channel independently, and reveal the nodal structures:
\begin{itemize}
    \item $d_{x,0}$ and/or $d_{y,0}$: The pairing vanishes at $\sin 2\theta = 0$ or $\sin \phi =0$. This leads to two intersected nodal loops, one at $k_z=0$ ($\theta = \frac{\pi}{2}$) and the other at $k_x=0$ ($\phi=0,\pi$). 
    \item $d_{x,z}$ and/or $d_{y,z}$: Two intersected nodal loops, one at $k_z=0$ and the other at $k_y=0$. 
    \item $d_{x,x}$ and/or $d_{y,x}$: Two parallel nodal loops at $\theta = \frac{\pi}{4}$ and $\theta=\frac{3\pi}{4}$. 
    \item $d_{x,y}$ and $d_{y,y}$: No apparent nodal structure. 
\end{itemize}
Since these nodal structures are completely determined by the information of wavefunctions of normal states and independent of the detials of $d_{i,j}$, we generally refer to them as geometry obstructed nodal pairing. The topologically obstructed nodal pairing is a special case of the geometry obstructed nodal pairing, where the existence of nodal points can be attributed to some well-defined topological invariants of the normal states. Importantly, for Dirac semimetals, $d_{x/y,0}, d_{x/y,z}$ channels are always geometrically-obstructed nodal at poles ($\theta=0,\pi$) and the equator ($\theta=\pi/2$).

\subsection{Symmetry analysis}
Next, let us analyze constraints imposed by symmetries on $d_{ij}$. Following Eq.~\eqref{eq:rep}, we identify different pairing channels as different irreps of $C_{4h}\times \mathcal{T}$ group. For simplicity, we label each irrep by $(X_{g/u},\alpha_{\mathcal{T}})$, where $X_{g/u}$ specify the irrep of $C_{4h}$. The symmetry operators act on $\tau_{i}\sigma_{j}$ as

\begin{equation}
\label{eq:symmetryDD}
\begin{aligned}
& C_{4}\tau_{x/y}\sigma_{x,y}C_{4}^{T}=s_{x/y}\sigma_{x,y}, \ C_{4}\tau_{x/y}\sigma_{\pm}C_{4}^{T}=\mp i s_{x/y}\sigma_{\pm}
\\
& U_{\mathcal{T}} (\tau_{x/y}\sigma_{0})^{\star} U_{\mathcal{T}}^{T}=s_{x/y}\sigma_{0},\ U_{\mathcal{T}} (\tau_{x/y}\sigma_{x,y,z})^{\star}U_{\mathcal{T}}^{T}=-\tau_{x/y}\sigma_{x,y,z},
\end{aligned}
\end{equation}
where $U_{\mathcal{T}}=i\tau_{x}\sigma_{y}$ is the unitary part of the $\mathcal{T}$. Let us now concentrate on the nodal structures at the poles and the equator of the FS, which are most relevant to the discussion in the main text.  Since $d_{x/y,0/z}$ are already guaranteed to be nodal at poles and equators by the geometry of normal states, here we list the symmetry constraints for $d_{x/y,x/y}$ in different irreps:

\begin{itemize}
    \item ($X_g$,+1) with $X=A,B,^{1}E,^{2}E$: First, $d_{x,x}$ and $d_{y,y}$ are ruled out, because Fermi statistics requires them to be odd in $\mathbf{k}$ but spatial inversion symmetry requires them to be even in $\mathbf{k}$. Next, $d_{x,y}$ and $d_{y,x}$ are purely imaginary because of $\mathcal{T}$. Finally, if $X= ^{1}E,^{2}E$, $d_{x,y}$ and $d_{y,x}$ are nontrivial irrep of $C_{4}$ with characters $\pm i$, and thus are odd in both $(k_{x},k_{y})$ and $k_{z}$. This indicates that $d_{x,y}$ and $d_{y,x}$ vanish both at poles and the equator.  For $X=B$, $d_{x,y}$ and $d_{y,x}$ are nontrivial irrep of $C_{4}$ with characters $\pm i$, and thus must vanish at the poles that are $C_{4}$ fixed point. For $X=A$, there is no further constraints, and the system can be gapped out by a $s$-wave pairing.
    \item ($X_g$,-1) with $X=A,B,^{1}E,^{2}E$: Similar to the first case, except that $d_{x,y}$ and $d_{y,x}$ are enforced to be purely real by $\mathcal{T}$ in this case.
    \item ($X_{u}$,+1) with $X=A,B,^{1}E,^{2}E$: All anti-symmetric $\tau_{i}\sigma_{j}$s are ruled out by the spatial inversion symmetry, and the only available terms are $d_{x,x}$ and $d_{y,y}$, which are enforced to be purely real by $\mathcal{T}$. For $X=^{1}E,^{2}E$, $d_{x,x}$ and $d_{y,y}$ are odd in $(k_{x},k_{y})$ but even in $k_{z}$, and thus will vanish at poles. For $X=B$,  $d_{x,x}$ and $d_{y,y}$ are odd in $k_{z}$ and form nontrivial irrep of $C_{4}$, and thus they vanish at both the poles and the equator.  For $X=A$, $d_{x,x}$ and $d_{y,y}$ are odd in $k_{z}$ and thus vanish at the equator.
    \item ($X_{u}$,-1) with $X=A,B,^{1}E,^{2}E$: Similar  to the third case, except that $d_{x,y}$ and $d_{y,x}$ are enforced to be purely imaginary by $\mathcal{T}$.
\end{itemize}
The s-wave and p-wave pairing in Eq.~(9) of the main text correspond to $(A_{g},+1)$ and $(B_{u},+1)$, but they fall in the same $A_{1}$ irrep of $T_{d}$ group.

\begin{figure}[h]
    \centering
    \includegraphics[width=0.5\columnwidth]{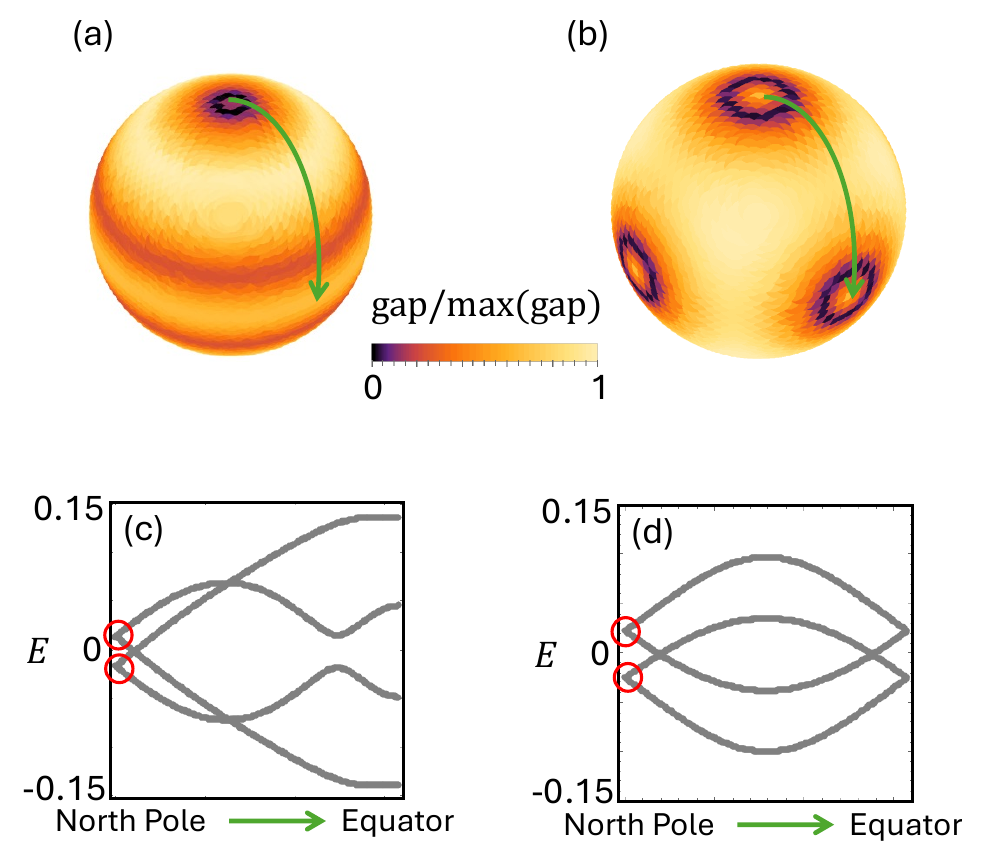}
    \caption{(a-b) Superconducting gap over the FS for a Dirac-dipole semimetal with the p-wave pairing and a Luttinger semimetal with both s-wave and p-wave pairings, respectively. (c-d) Energy spectra along the green-arrow paths in (a-b).  The calculations are done for $\mu=-0.2$, $t=0.2$, $\Delta_{s}=0.01$, and $\Delta_{p}=0.05$. The red circles highlight the degeneracy rooted in the dipole-obstructed intra-sector pairing. Note that the surfaces in (a-b) are FSs with an approximately (but not exactly) constant energy, which are specifically chosen to better illustrate the nodal structures of projected pairings.}
    \label{fig:SOC}
\end{figure}

\section{Perturbations from asymmetric spin-orbit coupling}

In real materials, the spatial inversion symmetry breaking spin-3/2 pairing happens because of the asymmetric spin-orbit coupling (ASOC)~\cite{pairing2016brydon,kim2018beyond,tuning2021ishihara}. Here, we consider the perturbative effect of the ASOC on the dipole-obstructed nodal points.  Specifically, we consider the linear Dresselhaus term
\begin{equation}
h_{SOC}=t (k_x (J_yJ_xJ_y-J_zJ_xJ_z)+ 
k_y (J_zJ_yJ_z-J_xJ_yJ_z)+k_z (J_xJ_zJ_x-J_yJ_zJ_x),
\end{equation}
where $J_{x,y,z}$ are the rotation generators for $j=3/2$:

\begin{equation}
    J_x=\frac{1}{2}\left(\begin{array}{cccc}
0 & \sqrt{3} & 0 & 0 \\
\sqrt{3} & 0 & 2 & 0 \\
0 & 2 & 0 & \sqrt{3} \\
0 & 0 & \sqrt{3} & 0
\end{array}\right), J_y=\frac{1}{2}\left(\begin{array}{cccc}
0 & -\sqrt{3} i & 0 & 0 \\
\sqrt{3} i & 0 & -2 i & 0 \\
0 & 2 i & 0 & -\sqrt{3} i \\
0 & 0 & \sqrt{3} i & 0
\end{array}\right), J_z=\frac{1}{2}\left(\begin{array}{cccc}
3 & 0 & 0 & 0 \\
0 & 1 & 0 & 0 \\
0 & 0 & -1 & 0 \\
0 & 0 & 0 & -3
\end{array}\right).
\end{equation}

We add $h_{SOC}$ to the Luttinger-Khon model and study its effect on the dipole-obstructed intra-sector pairing, which manifests as degeneracy highlighted by the purple points in Fig.~3(f) of the main text. We first look at the Dirac-dipole semimetal with $p$-wave pairing. As shown in Fig.~\ref{fig:SOC}(a) and (c), the nodal points due to the dipole-obstructed intra-FS pairing at two poles are deformed into nodal loops. Therefore, the effect of the ASOC resembles that of the s-wave pairing term, which couples the nodal points in different FSs and partially lifts the four-fold degeneracy guaranteed by the dipole-obstructed intra-pairing. By turning on both $\alpha$ and $\Delta_{s}$, we find that the nodal loop is still robust with only a slight change to its location on the FS, as shown in Fig.~\ref{fig:SOC}(b) and (d).

\begin{figure}[h]
    \centering
    \includegraphics[width=0.6\columnwidth]{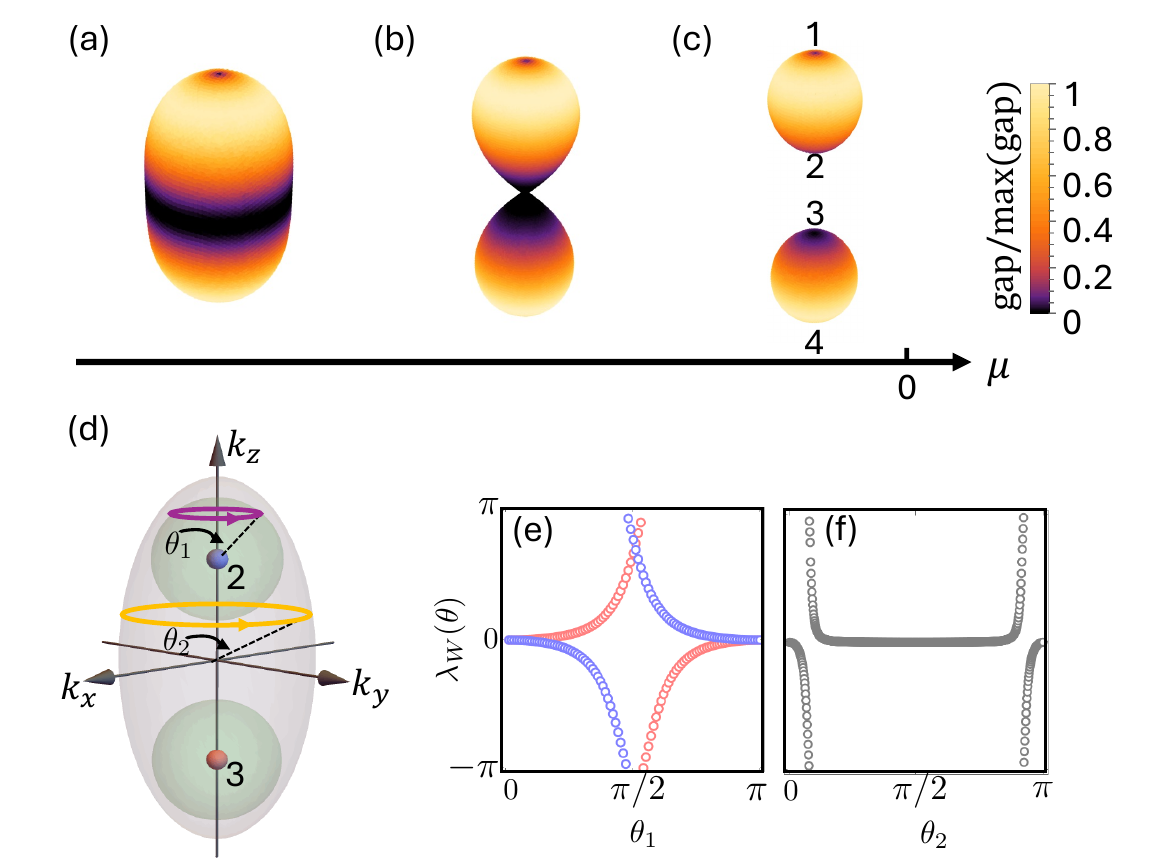}
    \caption{(a)-(c) Superconducting gap over FS for Berry-dipole WSM ($\Sigma=0.1$) at various $\mu=-0.2,-0.1,-0.05$. The numbers in (c) label the nodal points. (d) Illustration of Gaussian surfaces enclosing nodal points 2 (blue) and 3 (red). (e) Berry phase versus polar angle over the green spheres in (c), where the blue (red) circles are calculated for the blue (red) nodal point in (c). (f) Berry phase versus polar angle over the gray ellipsoid in (c).}
    \label{fig:BDSC}
\end{figure}

\section{Berry-dipole superconductor}
In this part, we discuss a case where the dipole-obstructed BdG nodal points themselves carry a Berry-dipole charge. This case is a BdG analog of the Berry-dipole semimetals~\cite{Sun2018conversion,Nelson2022delicate}, and thus is referred to as a Berry-dipole superconductor.

Consider the BdG Hamiltonain in Eq.~(4) in the main text, and choose $\Sigma=0.1$ and $\Delta(\mathbf{k})=k_{z}\sigma_{x}$. The normal states exhibit Weyl points, and thus when $|\mu|<|\Sigma|$, the FSs are two spheres as shown in Fig.~\ref{fig:BDSC}(c). Focusing on nodal points $2$ and $3$, numerical calculations of Berry phases along the azimuthal direction over Green and Gray Gaussian surfaces shown in Fig.~\ref{fig:BDSC}(d) reveal that (i) points $2$ and $3$ are BdG Weyl nodes [c.f. Fig,~\ref{fig:BDSC}(e)]; and (ii) they carry a Berry-dipole charge [c.f. Fig,~\ref{fig:BDSC}(f)], i.e., on a Gaussian surface surrounding both of the nodes, there is a $2\pi$ ($-2\pi$) quantized Berry flux over the upper (lower) hemi-sphere. This Berry-dipole charge clearly puts constraints on pair annihilation, and thus when nodal points $2$ and $3$ meet each as we vary the chemical potential, they will become a nodal loop in $k_{z}=0$ plane instead of being gapped out. The Berry dipole charge presents here because $\Delta(\mathbf{k})=0$ at $k_{z}=0$, and hence all eigenstates of the BdG Hamiltonian are identical on the $k_{z}=0$ plane and thus over the equator of the Gaussian surface surrounding both nodes. Then, the equator of this Gaussian surface can always be viewed as a point, and each half of this Gaussian surface is effectively closed. Consequently, the Berry flux will be quantized over each half of the Gaussian surface. 

We note that this Berry-dipole superconductor is {\it not} a symmetry-protected crystalline superconducting phase, because we can include symmetric $d_{\pm}\sigma_{\pm}$ pairings to make the eigenstates over the equator momentum-dependent. The $C_{4h}$ symmetry considered in our work cannot isolate $k_{z}\sigma_{x}$ pairing and rule out all others.  Thus, it could be an interesting direction to search for symmetry-protected Berry-dipole superconductors and conduct relevant classification studies.

\bibliography{refs}